\documentstyle[aps,preprint,tighten,epsfig]{revtex}
\begin{document}
\draft
\def\etal{{\em et\ al.\/}, }
\def\eg{{\em e.g.\/}, }
\def\ie{{\em i.e.\/}, }
\def\beq{\begin{equation}}
\def\eeq{\end{equation}}
\def\beqarr{\begin{eqnarray}}
\def\eeqarr{\end{eqnarray}}
\def\ix{\int_{-L/2}^{L/2}dx\,}
\def\ixi{\int dx\,}
\def\ii{\int_{-\infty}^{\infty}}
\def\eps{\varepsilon}
\def\ov{\over}
\def\ol{\overline}
\def\sgn{{\rm sgn}}
\def\ph{\hat{\phi}}
\def\Pt{\tilde{\Psi}}
\def\tr{{\rm tr}}
\def\det{{\rm det}}
\def\la{\langle}
\def\ra{\rangle}
\def\8{\infty}
\def\one{{\bf 1}}
\def\lp{\lambda'}
\def\Zint{
  {\mathchoice
    {{\sf Z\hskip-0.45em{}Z}}
    {{\sf Z\hskip-0.44em{}Z}}
    {{\sf Z\hskip-0.34em{}Z}}
    {{\sf Z\hskip-0.35em{}Z}}
    }
  }
\def\del{\partial}
\def\delx{\partial_x}
\def\delxp{\partial_{x'}}
\def\delt{\partial_{\tau}}
\def\delT{\partial_t}
\def\XLL{{$\chi$LL}}
\def\gtlt{\,{\stackrel {\scriptscriptstyle >}{\scriptscriptstyle <}}\,}
\def\bh{{\hat \beta}}
\def\ph{{\hat \phi}}
\def\XSG{{$\chi$SG }}
\def\bJ{{\bf J}}
\def\Lt{L^{(1/2)}}
\def\Lh{L'^{(1/2)}}
\def\lh{{\hat \lambda}}
\def\nb{{\bar \nu}}
\def\tb{{\bar \tau}}
\def\ID{I}
\def\Ab#1{{\bf A}^{(#1)}}
\def\Bb#1{{\bf B}^{(#1)}}
\def\Cb#1{{\bf C}^{(#1)}}
\def\Af#1{\tilde{\bf A}^{(#1)}}
\def\Bf{\tilde{\bf B}}
\def\Cf#1{\tilde{\bf C}^{(#1)}}
\def\Rf#1{\tilde{\bf R}^{(#1)}}
\def\Gf#1{\tilde{\bf G}^{(#1)}}
\def\X#1{2\pi(x-v_{#1}t+i\epsilon_t)}
\def\XP#1{2\pi(x-x'-v_{#1}t+i\epsilon_t)}
\def\Tt{{\tilde T}}
\def\Lt{{\tilde L}}
\def\mat{\sf}
\def\Wt{{\widetilde W}}

\title{Notes on Infinite Layer Quantum Hall Systems}

\author{J. D. Naud$^1$, Leonid P. Pryadko$^2$, and S. L. Sondhi$^1$.}

\address{
$^1$Department of Physics,
Princeton University,
Princeton, NJ 08544, USA}

\address{
$^2$School of Natural Sciences,
Institute for Advanced Study,
Olden Lane,
Princeton, NJ 08540, USA}

\date{\today}
\maketitle
\begin{abstract}
We study the fractional quantum Hall effect in three dimensional
systems consisting of infinitely many stacked two dimensional 
electron gases placed in transverse magnetic fields. This limit
introduces new features into the bulk physics such as quasiparticles
with non-trivial internal structure, irrational braiding phases,
and the necessity of a boundary hierarchy construction for 
interlayer correlated states. The bulk states host a family of
surface phases obtained by hybridizing the edge states in each
layer. We analyze the surface conduction in these phases by means of 
sum rule and renormalization group arguments and by explicit 
computations at weak tunneling in the presence of disorder. We find
that in cases where the interlayer electron tunneling is not relevant
in the clean limit, the surface phases are chiral semi-metals that
conduct only in the presence of disorder or at finite temperature.
We show that this class of problems which are naturally formulated
as interacting bosonic theories can be fermionized by a general
technique that could prove useful in the solution of such ``one
and a half'' dimensional problems. 
\end{abstract}
\pacs{}

\section{Introduction}
\label{sec:intro}

The quantum Hall effect arises from physics that is special to two
dimensions.  Nevertheless, it was demonstrated in an early
experiment\cite{stormer1} that it survives a small amount of three
dimensionality; more precisely, that an infinite layer system with
interlayer tunneling weak compared to the single layer (mobility) gap
would continue to exhibit dissipationless transport and a quantized
Hall resistance per layer. The quantized Hall phases in the organics
also rely on the same effect\cite{organics}.  In the presence of
tunneling the chiral edge states which exist in each layer hybridize
and yield a family of ``one and a half'' dimensional phases that live
on the surfaces of the three dimensional systems and exhibit
interesting transport in the direction transverse to the layers.

Following the pioneering theoretical
work\cite{Chalker-Dohmen,Balents-Fisher}, a flurry of
work\cite{Betouras-Chalker} has focused on the integer Hall effect in
infinite layers systems with a particular emphasis on the properties
of the chiral metal that forms at their surface in the quantum Hall
phases and some of this has found support in experiments\cite{druist}.
Recently, interaction effects and magnetic order at $\nu=1$ (per
layer) have been discussed as well\cite{brey}. All of this work has
antecedents in work on multi-component systems such as double layer
devices\cite{gm1}, but the limit of infinitely many layers sharpens
quantitative differences into qualitatively new features, \eg a
different universality class for the transitions out of the quantum
Hall phases\cite{Chalker-Dohmen}.

In this paper we extend the study of three dimensional quantum Hall phases
in two directions. First, we offer a systematic analysis of the simplest
{\it fractional} quantum Hall phases in infinite layer systems with a
special focus on the phases that exhibit interlayer correlations. Here
we find several novel bulk properties, most notably a non-trivial structure
for the quasiparticles. This part of our work builds on the pioneering and 
early work of Qiu, Joynt and MacDonald (QJM)\cite{QJM}, who studied the 
energetics of multilayer states, and noted the irrational partitioning
of the quasiparticle charge. The second axis of our work is the analysis
of the ``one and a half'' dimensional phases that arise at the surfaces
of fractional quantum Hall phases. In this regard we report a general
analysis of the surface conduction by combining renormalization group
arguments (which have strong consequences for the ground state structure
in chiral systems) and a conductivity sum rule. We also carry out 
computations of the surface conductivity of the disordered system in
a weak tunneling expansion for a large class of states. Together, these
show that the surfaces of states with marginal or irrelevant electron
tunneling are semi-metallic, in that they are perfect insulators at
all frequencies in the clean limit but begin to conduct at finite
temperature and disorder.  Some of the results derived here were
announced previously in a companion paper\cite{nps3}.

This last theme, of analyzing the surface phases, is interesting from
a more theoretical standpoint in that the surface phases are ``half''
a dimension up from one dimension where exact solutions are often
possible. We have not made very much progress on this front. What we
have accomplished is to solve the case of bilayers in some generality
\cite{nps1,nps2}, find special solutions for three and four layers,
and recast the infinite layer problems in algebraic and non-standard
fermionized formulations that appear potentially fruitful. We report
these here in the hope that some readers will find them useful in
carrying this program to completion.  We should also note that the
construction of more complex bulk states than those considered in this
paper could well lead to a richer class of such problems.

The outline of this paper is as follows.  We begin in
Section~\ref{sec:bulk} with a discussion of the bulk properties of
multilayer fractional states, in particular the novel features of
their quasihole excitations which include a non-trivial internal
charge structure and irrational statistics.  In Section~\ref{sec:edge}
we discuss the edge theory of general multilayer states in the
presence of nearest-neighbor single-electron tunneling.  We specialize
to the case where the tunneling is marginal in
Section~\ref{sec:marginal}, first considering the clean limit and then
adding disorder and interactions.  In Section~\ref{sec:irrelevant} we
extend our analysis to states where tunneling is irrelevant.  We
conclude in Section~\ref{sec:fermionizations} by presenting a method
for fermionizing the multilayer edge theory via the addition of
auxiliary degrees of freedom.  The appendices contain a discussion of
an exactly soluble model problem with marginal tunneling
(App.~\ref{sec:mmmm}), exact solutions for the spectra of two edge
theories for the special case of four layers
(App.~\ref{sec:exactN=4}), and some technical details including Klein
factors (App.~\ref{sec:Kleins}), and the proof of an assertion made in
the main text (App.~\ref{sec:converse}).

\section{Bulk Properties}
\label{sec:bulk}

Consider a system which consists of $N$ parallel layers of 2DEGs in a
strong perpendicular magnetic field.  We assume there is a
confining potential which restricts the electrons in each layer to a
region with the topology of a disc.  The phase diagram of this system
was first investigated by Qiu, Joynt, and MacDonald (QJM)
\cite{QJM}.  At low electron densities they found a variety
of solid phases, while at higher densities they found the ground state
to be an incompressible fluid described by the $N$-layer
generalization of the Laughlin wavefunction:
\beq
\Psi_{0}(\{z_{i,\alpha}\})=\prod_{i=1}^N\prod_{\alpha < \beta}^{N_i}
(z_{i\alpha}-z_{i\beta})^{K_{ii}}
\prod_{i<j}^N\prod_{\alpha=1}^{N_i}\prod_{\beta=1}^{N_j}
(z_{i\alpha}-z_{j\beta})^{K_{ij}}
e^{-\sum_{\alpha,i}|z_{\alpha,i}|^2/4}\label{eq:multilaugh}.
\eeq
Here $z_{i\alpha}$ is the coordinate of electron $\alpha$ in layer
$i$, and $N_i$ is the number of electrons in layer $i$.  The exponents
are specified by a symmetric, $N\times N$ matrix $K$.  The diagonal
elements of this matrix determine the electron correlations within
each layer and the off-diagonal elements specify the correlations
between layers.  For the wavefunction to describe electrons the
diagonal elements of $K$ must be odd integers.  The filling factor in
layer $i$ is given by
\beq
\label{eq:multinu}
\nu_i=\sum_{j=1}^N(K^{-1})_{ij}.
\eeq
Note that it is possible to construct more complicated multilayer
states by beginning with the states in Eq.~(\ref{eq:multilaugh}) and
carrying out a generalization of the single-layer Haldane-Halperin
hierarchy construction\cite{greiter-mcdonald,chen}.  For example, at
the first level of the hierarchy the effective $K$ matrix is given by
\beq
\label{eq:K1}
K_{(1)}=K+AP^{-1},
\eeq
where $A$ is a diagonal $N\times N$ matrix with elements $\pm 1$, and
$P$ is a symmetric $N\times N$ matrix with even integer elements along
the diagonal and integer elements everywhere else.  The sign of
the element $A_{jj}$ determines whether the excitations in layer $j$ are
quasiholes (+1) or quasielectrons ($-1$) and the matrix $P$ specifies
the (bosonic) multilayer state into which these excitations condense.
Using the matrix $K_{(1)}$ in Eq.~(\ref{eq:multinu}) gives the filling
factors at the first level of the hierarchy.

We will restrict ourselves to the case where the matrix $K$ is
tridiagonal
\beq
\label{eq:Ktri}
K_{ij}=m\delta_{ij}+n(\delta_{i,j-1}+\delta_{i,j+1}),
\eeq
where $m$ and $n$ are nonnegative integers.  Since we are interested
in the limit of large $N$, we will impose periodic boundary conditions
along the direction perpendicular to the layers by the identification
$N+1\equiv 1$.  One of the difficulties encountered when working
without periodic boundary conditions is discussed briefly below.  
The standard convention is to refer to the state whose
$K$ matrix is of the form (\ref{eq:Ktri}) as the ``$nmn$'' state.  
Using Eqns.~(\ref{eq:multinu}) and
(\ref{eq:Ktri}), the filling factor per layer in the $nmn$ state is
\beq
\label{eq:nu_nmn}
\nu={1\ov m+2n}.
\eeq
From this expression we see that there are multiple states with the
same filling factor per layer.  For example, the states 050 and 131
both have filling factor $\nu=1/5$ per layer.  QJM found that as the
interlayer separation $d$ was decreased there is a phase transition
between the 050 state in which there are no interlayer correlations
and the 131 state in which electrons in neighboring layers are
correlated.

Note that Eq.~(\ref{eq:nu_nmn}) applies to the case $N\rightarrow \8$ or
to the case of finite $N$ with periodic boundary conditions.  For a
physically realizable multilayer system, \ie one with $N$ finite but
without periodic boundary conditions, one finds that the filling factor
$\nu_j$ is not independent of the layer index $j$.  If one solves
Eq.~(\ref{eq:multinu}) for finite $N$ without periodic boundary conditions,
using the $K$ matrix given in Eq.~(\ref{eq:Ktri}) one finds
\beq
\label{eq:nonuninu}
\nu_j={1\ov m+2n}\left[{1-{y^{j}+y^{N+1-j}\ov 1+y^{N+1}}}\right],
\eeq
where
\beq
\label{eq:y}
y=-{m\ov 2n}+\sqrt{\left({m\ov 2n}\right)^2-1}.
\eeq
From this form we find that for large $N$ the filling factor near the
middle of the multilayer stack is given approximately by
Eq.~(\ref{eq:nu_nmn}), but exhibits oscillations about this value as
one approaches the end layers.  Such non-uniformities in the electron
density would cost electrostatic energy, and could be mitigated by the
formation of quasihole excitations near the outermost layers.
Therefore, one would expect that the true ground state for a finite
multilayer with interlayer correlations would be determined by
balancing the creation energy of quasiholes against the electrostatic
energy of the density oscillations.  It is interesting to note that to
construct a state in a finite stack with uniform filling by carrying
out the hierarchy construction in only the outermost layers one must go
to an infinite level in the hierarchy.  Henceforth we will avoid these
complications by working with periodic boundary conditions, arguing
that in the limit of large $N$ they can safely be ignored.

\subsection{Quasihole Excitations}
\label{ss:quasiholes}

In the last section we learned that for a range of electron densities
the ground state of the multilayer system is an incompressible state
described by the wavefunction (\ref{eq:multilaugh}).  It is natural to
ask about the properties of the excitations of this state.  By analogy
with the single-layer quantum Hall effect we write the wavefunction
for a single quasihole at point $\xi$ in layer $j$ as:
\beq
\label{eq:qhole}
\Psi_{(j,\xi)}(\{z_{i\alpha}\})=\prod_{\alpha=1}^{N_j}(z_{j\alpha}-\xi)
\Psi_0(\{z_{i\alpha}\}).
\eeq
If one interprets $|\Psi_{(j,\xi)}|^2\equiv e^{-\beta V(\{z\})}$ as
the Boltzmann factor for a classical 2D generalized Coulomb plasma at
inverse temperature $\beta$ with an impurity at $\xi$, the perfect
screening conditions give:
\beq
\label{eq:perfectscreen}
K_{ik}Q^{(j)}_k=\delta_{ij},
\eeq
where $Q^{(j)}_k=(K^{-1})_{kj}$ is the local deviation of the charge in
layer $k$ from its ground state value, due to the quasihole in layer
$j$\cite{QJM}.  The total charge of the quasihole is
\beq
\label{eq:totalcharge}
Q=\sum_k Q^{(j)}_k=\sum_k (K^{-1})_{jk}=\nu_j={1\ov m+2n},
\eeq
where we have used Eq.~(\ref{eq:multinu}).  Thus, the relation between
the total quasihole charge and the filling factor is the same as in
the single-layer quantum Hall effect.  However, if there are
interlayer correlations, \ie $n\neq 0$, the quasihole charge is
spread over many layers.  Indeed, in the limit $N\rightarrow \8$ the
individual charges are {\em irrational}:
\beq
\label{eq:indcharges}
Q^{(j)}_k={1\ov\sqrt{m^2-4n^2}} \left({\sqrt{m^2-4n^2}-m\ov 2n}\right)^{|k-j|}.
\eeq

The same Coulomb plasma analogy, combined with the mean field
approximation, allows us to find the spatial distribution of the extra
charge excited by the hole.  The Debye screening equation for an
average ``potential'' $V_i^{(j,\xi)}(z)$ acting on an electron at the
point $z$ in layer $i$ can be written as
\begin{equation}
  \label{eq:mf-plasma-debye}
  -\beta\nabla^2V_i^{(j,\xi)}=4\pi\delta_{ij}\delta(z-\xi)-2
   +4\pi\sum_k K_{ik}\,n_k,   
\end{equation}
where the single-particle average charge density
\begin{equation}
  \label{eq:mf-plasma-charge-density}
  n_i=\bigl(2\pi {\textstyle\sum_j}
  K_{ij}\bigr)^{-1}\,e^{-\beta V_i}
\end{equation}
is chosen so that at $V_i=0$ the unperturbed density would be
restored.  Clearly, the screening in each successive layer is limited
by the Debye screening length which is on the order of the magnetic length
($l=1$ in chosen units).  Therefore, as the amount of charge induced
by the hole is reduced with the separation along the stack, this
charge also spreads over a wider and wider area, further reducing
the charge density.  The shape of the induced charge distribution can
be found by linearizing the
density~(\ref{eq:mf-plasma-charge-density}) in
Eq.~(\ref{eq:mf-plasma-debye}) and solving the resulting set of linear
partial differential equations by Fourier transformation.  In
particular, for the r.m.s.\ spread of the distribution created in
layer $j$ by the quasihole in layer $i$ we obtain 
\begin{equation}
  \label{eq:mf-plasma-charge-spread}
  \left\langle r^2_{|i-j|}\right\rangle=-2
  I_2(|j-i|)/I_1(|j-i|),\quad{\rm where}\quad
  I_p(j)\equiv \int_0^{2\pi} dq\,{\cos(q j)\over K^p_q},
\end{equation}
and $K_q=m+2n\cos q$ is the Fourier transform of Eq.~(\ref{eq:Ktri}).
Specifically, for the 131 state we find $\langle
r^2_j\rangle = {4\,\left( 3 + {\sqrt{5}}\,|j| \right) }/{5} $.  If we
were to color in the perturbed charge density, the hole
would produce a characteristic ``hourglass'' shape.

To compute the statistics of these excitations we follow the method
of Arovas, Schrieffer, and Wilczek\cite{ASW} and consider a state with two
quasiholes, one at point $\xi$ in layer $j$ and one at point $\eta$ in
layer $k$:
\beq
\label{eq:twoholes}
\Psi_{(j,\xi;k,\eta)}=\prod_{\alpha=1}^{N_j}(z_{j\alpha}-\xi)
\prod_{\beta=1}^{N_k}(z_{k\beta}-\eta)\Psi_0.
\eeq
The Berry phase for moving the quasihole at $(k,\eta)$ around a
closed loop of radius $R$ containing the quasihole at $(j,\xi)$ is
\beqarr
\gamma_B&=&i\oint_{\xi}d\eta\, \la \Psi_{(j,\xi;k,\eta)}|\del_{\eta}
\Psi_{(j,\xi;k,\eta)}\ra\nonumber \\
&=&i\oint_{\xi}d\eta\,\int d^2z\,{1\ov \eta-z}
\la\Psi_{(j,\xi;k,\eta)}|\sum_{\beta=1}^{N_k}\delta^{(2)}(z-z_{k\beta})
|\Psi_{(j,\xi;k,\eta)}\ra.\label{eq:berrys}
\eeqarr
The expectation value appearing in this equation is just the electron
density at the point $z$ in layer $k$ for the two-quasihole state.  We
can write this as
\beq
\la\Psi_{(j,\xi;k,\eta)}|\sum_{\beta=1}^{N_k}\delta^{(2)}(z-z_{k\beta})
|\Psi_{(j,\xi;k,\eta)}\ra=
\rho_k^{(0)}(z)+\delta\rho_k^{(k,\eta)}(z)
+\delta\rho_k^{(j,\xi)}(z),
\label{eq:rho2qh}
\eeq
where $\rho_k^{(0)}(z)$ is the density in the ground state,
$\delta\rho_k^{(k,\eta)}(z)$ is the change in density due to the
quasihole at $(k,\eta)$ and $\delta\rho_k^{(j,\xi)}(z)$ is the change
in density due to the quasihole at $(j,\xi)$.  If we substitute
Eq.~(\ref{eq:rho2qh}) into Eq.~(\ref{eq:berrys}), the $\rho_k^{(0)}$
term gives the Aharonov-Bohm phase, the $\delta\rho_k^{(k,\eta)}$
term gives zero by symmetry, and hence 
\beq
\label{eq:berry2}
\gamma_B^{({\rm stat})}=i\int d^2z\,\delta\rho_k^{(j,\xi)}(z)\oint_{\xi}d\eta\,
{1\ov \eta-z}=-2\pi Q^{(j)}_k,
\eeq
where once again $Q^{(j)}_k$ is the charge deviation in layer $k$ due
to a quasihole in layer $j$, which we found above to be equal to
$(K^{-1})_{kj}$, see Eq.~(\ref{eq:perfectscreen}).  The statistical
angle for interchanging two quasiholes in the same layer is
\beq
\label{eq:statangle}
\Delta\gamma={1\ov 2}\gamma_B^{({\rm stat})}= 
-\pi (K^{-1})_{jj}=-{\pi\ov\sqrt{m^2-4n^2}}.
\eeq
We find the statistical angle of the quasiholes is an {\em irrational}
multiple of $\pi$ in the infinite layer system.

\section{Edge Theory}
\label{sec:edge}

In this section we consider the edge theory of the multilayer $nmn$
state in the presence of interlayer single-electron tunneling.  The
low-energy effective Hamiltonian of the edge theory in the absence of
tunneling is\cite{wen}:
\beq
\label{eq:multiH0}
  {\cal H}_0=\ix {1\over{4\pi}}V_{ij} 
  \,:\!\delx u_i\,\delx u_j\!:,
\eeq
where $V$ is a symmetric, positive definite, $N\times N$ matrix which
depends on the interactions and confining potentials at the edge, and
we take the $x$-axis along the edges of length $L$.  The $N$ chiral
bosons appearing in this Hamiltonian obey the equal-time commutation
relations
\beq
\label{eq:multiCR}
[u_i(x), u_j(x')]=i\pi K_{ij}\,{\rm sgn}(x-x').
\eeq
The electron charge density and the electron creation operator in
layer $i$ are given by
\beq
\label{eq:multirho,psi}
\rho_i(x)={1\over 2\pi}(K^{-1})_{ij}\delx u_j(x),\quad
\Psi_{i}^{\dagger}(x)\propto e^{-i u_i(x)}.
\eeq

Note that because of the factor of $K^{-1}$ in the expression for the
density operator (\ref{eq:multirho,psi}) there is a non-trivial
relation between the matrix $V$ appearing in the Hamiltonian and the
matrix determining the density-density couplings, which we will call
$U$.  If we rewrite the Hamiltonian (\ref{eq:multiH0}) as
\beq
\label{eq:HfromT}
  {\cal H}_0=\ix U_{ij}\,:\!\rho_i(x)\rho_j(x)\!:,
\eeq
which serves as our definition of $U$, we find by equating these two
forms that
\beq
\label{eq:T-V}
V={1\ov\pi}K^{-1}UK^{-1}.
\eeq

We assume that the system is translationally invariant along the
direction perpendicular to the layers, which we shall take to be the
$z$-axis.  We shall often refer to this as the ``vertical'' direction.
Specifically, we assume the matrices $K_{ij}$ and $U_{ij}$ depend only
on the difference $|i-j|$.  The tridiagonal form of $K$ given above in
Eq.~(\ref{eq:Ktri}) certainly obeys this constraint, provided we recall
that we are assuming periodic boundary conditions along $z$.  From
Eq.~(\ref{eq:T-V}) we see that if both $K$ and $U$ are translationally
invariant, so is $V$.  Hence to diagonalize $K$ and $V$ we perform a
discrete Fourier transformation in the $z$ direction, defining:
\beq
\label{eq:thetaq}
u_q(x)={1\ov\sqrt{N}}\sum_{j=1}^{N}e^{-iqj}u_j(x),
\eeq
where 
\beq
\label{eq:qset}
q\in\left\{{2\pi n/N |n=0,1,\ldots ,N-1}\right\},
\eeq
and we identify $q$ and $q+2\pi k$ for any integer $k$.

Using transformation (\ref{eq:thetaq}), the commutation relations
(\ref{eq:multiCR}) become
\beq
\label{eq:thetaqCR}
[u_q(x),u_{q'}(x')]=i\pi\delta_{q+q',0}K_q\sgn(x-x'),
\eeq
where
\beq
\label{eq:K_q}
K_q\equiv m+2n\cos q,
\eeq
are the eigenvalues of the matrix (\ref{eq:Ktri}).  
We will largely restrict our analysis to those states for which all
the edge modes move in the same direction, \ie the maximally chiral
case.  This requires that $K$ be positive definite.  From
Eqns.~(\ref{eq:qset}) and (\ref{eq:K_q}) this implies $n<m/2$.

Finally, we rescale the fields, defining
\beq
\label{eq:phiq}
\phi_q(x)={1\ov\sqrt{K_q}}u_q(x),
\eeq
which have conventionally normalized commutation relations
\beq
\label{eq:phiqCR}
[\phi_q(x),\phi_{q'}(x')]=i\pi\delta_{q+q',0}\sgn(x-x').
\eeq

Using the transformations (\ref{eq:thetaq}) and (\ref{eq:phiq}) to express the
Hamiltonian (\ref{eq:multiH0}) in terms of the fields $\phi_q$ we find
\beq
\label{eq:H0(phiq)}
  {\cal H}_0=\ix {1\over{4\pi}} v_q 
  \,:\!\delx\phi_q\delx\phi_{-q}\!:.
\eeq
The velocities $v_q$ are given by $v_q=V_qK_q$ where $V_q$ are
the eigenvalues of the $V$ matrix.  We now specialize to the case where
the density-density coupling matrix is of the form
\beq
\label{eq:Tpara}
U_{ij}=\pi v \delta_{ij}+\pi g(\delta_{i,j-1}+\delta_{i,j+1}),
\eeq
where $v$ is a velocity determined by the confining potential at the
edge and $g$ parameterizes the nearest-neighbor density-density
interactions between the layers.  Using the relation between $U$
and $V$ (\ref{eq:T-V}) we find that the eigenmode velocities
appearing in the Hamiltonian (\ref{eq:H0(phiq)}) are
\beq
\label{eq:v_q}
v_q={v+2g\cos q\ov m+ 2n\cos q}.
\eeq

Next we consider adding interlayer single-electron tunneling to the
multilayer edge theory.  The electron annihilation operator in layer
$i$ is given in Eq.~(\ref{eq:multirho,psi}) in terms of the original
bosonic field $u_i(x)$.  Using the transformations (\ref{eq:thetaq})
and (\ref{eq:phiq}) the tunneling operator between layers $j$ and
$j+1$ can be written
\beq
\label{eq:tunnelop}
\lambda \Psi_j(x)\Psi_{j+1}^{\dag}(x)+\,\,{\rm h.c.}\,\propto\,
\lambda e^{i\alpha_{jq}\phi_q(x)}+\,\,{\rm h.c.},
\eeq
where $\lambda$ is the tunneling amplitude and we have defined the
coefficients
\beq
\label{eq:alphas}
\alpha_{jq}\equiv{1\ov\sqrt{N}}e^{iqj}(1-e^{iq})\sqrt{K_q}.
\eeq
As defined in Eq.~(\ref{eq:multirho,psi}) the electron operators in
different layers do not obey proper anticommutation relations.  In
Appendix~\ref{sec:Kleins} we demonstrate that this can be remedied
without significantly altering the form of the tunneling operator
(\ref{eq:tunnelop}). 

The lowest-order perturbative RG flow of the tunneling amplitude
$\lambda$, assumed to be uniform along the edge, is controlled by the
scaling dimension, $\delta_{\lambda}$, of the tunneling operator
(\ref{eq:tunnelop}):
\beq
\label{eq:multiflow}
{d\lambda\ov d\ell}=\left({2-\delta_{\lambda}}\right)\lambda,
\eeq
where the short-distance cutoff increases as $\ell$ increases.
Since the fields $\phi_q$ obey canonically normalized commutation
relations, (\ref{eq:phiqCR}), we can read off the scaling dimension of the
tunneling operator from Eqns.~(\ref{eq:K_q}), (\ref{eq:tunnelop}), 
and (\ref{eq:alphas}):
\beq
\label{eq:deltalambda}
\delta_{\lambda}={1\ov 2}\sum_q \alpha_{jq}\alpha_{j-q}=m-n,
\eeq
where there is no sum on $j$.  Using Eqns.~(\ref{eq:multiflow})
and (\ref{eq:deltalambda}) we see that tunneling is relevant for
$m<n+2$, irrelevant for $m>n+2$, and marginal for $m=n+2$.

\begin{figure}[htbp]
  \begin{center}
    \leavevmode
    \epsfxsize=0.6\columnwidth
    \epsfbox{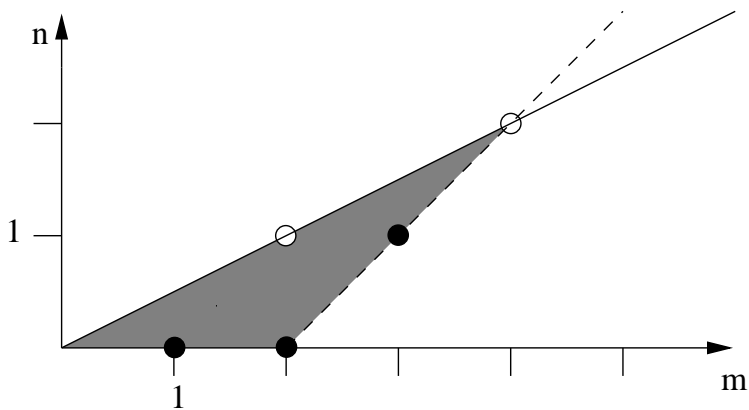}\vskip1mm
    \caption{
    The $m$-$n$ plane.  States to the right of the solid line are
    maximally chiral.  The dashed line separates states with
    irrelevant tunneling (to the right) from states with relevant
    tunneling (to the left).  The shaded region contains all
    maximally-chiral states with non-irrelevant tunneling: 010, 020,
    and 131 (dark circles).  The states 121 and 242 which lie on the
    boundary of the maximally-chiral region are also shown (open
    circles).}
    \label{fig:rel&stab}
  \end{center}
\end{figure}

If we combine this result with the condition of maximum chirality,
$n<m/2$, we obtain the diagram in Fig.~\ref{fig:rel&stab}.  The only
maximally-chiral state for which the tunneling is relevant is 010,
\ie uncorrelated $\nu=1$ layers.  There are two maximally-chiral
states for which tunneling is marginal: 131 and the bosonic state
020.  For all other maximally-chiral multilayer states interlayer
electron tunneling is irrelevant.

There are two bosonic states, 121 with relevant tunneling and 242 with
marginal tunneling, which lie on the boundary of the maximally-chiral
region.  The corresponding $K$ matrices have zero eigenvalues in the
limit of an infinite number of layers.  The complication this zero
eigenvalue adds is the possibility of ``soft'' modes which are not
strictly confined to the edge\cite{karlhede-sondhi}.  The 121 state is
the first of a sequence of ``pyramid'' states: 121, 12321,
1234321,$\ldots$, where in an obvious extension of the $nmn$ notation
the state $onmno$ has next-nearest-neighbor correlations specified by
the exponent $o$, in addition to nearest-neighbor ($n$) and intralayer
correlations ($m$), and similarly for the $ponmnop$ states,
etc\cite{fn-QJM}.  All of the pyramid states lie on the boundary of
the maximally chiral region (\ie their $K$ matrices are positive
semi-definite) and have relevant tunneling.  Were it not for the
complication from the soft modes, their edge theories would be exactly
soluble by fermionization as discussed in
Section~\ref{sec:fermionizations}.

The full Hamiltonian of the multilayer edge theory with tunneling can
be written using Eqns.~(\ref{eq:H0(phiq)}), (\ref{eq:tunnelop}), and
(\ref{eq:deltalambda}):
\beq
{\cal H}=\ix \left[{{1\over{4\pi}} v_q 
\,:\!\delx\phi_q\delx\phi_{-q}\!:
+\sum_j{\lambda\ov (2\pi a)^{\delta_{\lambda}}}
\left({e^{i\alpha_{jq}\phi_q(x)}+\,\,{\rm h.c.}}\right)}\right],
\label{eq:multiHfull}
\eeq 
where $a$ is a short distance cutoff.  In the following sections we
will be concerned with the $z$-axis conductivity of this theory, so we
will need the form of the current operator along $z$.  If
$\rho^{(2D)}_j(x,t)$ is the two-dimensional charge density operator in
layer $j$, the continuity equation can be written
\beq
\label{eq:continuity}
{\del\ov\del t}\rho^{(2D)}_j(x,t)={1\ov d}\left({{\cal J}^z_{j-1,j}(x,t)-
{\cal J}^z_{j,j+1}(x,t)}\right)-\delx{\cal J}^x_j(x,t),
\eeq
where ${\cal J}^x_j$ is the current density along the edge of layer $j$,
${\cal J}^z_{j,l}$ is the current density flowing from layer $j$ to layer
$l$, and $d$ is the layer separation.  Introducing the
discrete Fourier transforms
\beq
\label{eq:FTrho}
\rho^{(2D)}(k,q,t)=d\sum_j\int dx\,e^{-iqj}e^{-ikx}\rho^{(2D)}_j(x,t),
\eeq
with an analogous definition for ${\cal J}^x(k,q,t)$, and
\beq
\label{eq:FTJz}
{\cal J}^z(k,q,t)=d\sum_j\int dx\,e^{-iq(j-1/2)}e^{-ikx}{\cal J}^z_{j-1,j}(x,t),
\eeq
where the allowed values of the dimensionless $z$-momentum $q$ are
given in Eq.~(\ref{eq:qset}), the continuity equation
(\ref{eq:continuity}) becomes
\beq
\label{eq:cont2}
{\del\ov\del t}\rho^{(2D)}(k,q,t)=-{2i\ov d}\sin(q/2){\cal J}^z(k,q,t)
-ik{\cal J}^x(k,q,t).
\eeq
Note $\rho_j^{(2D)}(x,t)\equiv (1/d)\rho_j(x,t)$, where $\rho_j$ is given in
Eq.~(\ref{eq:multirho,psi}).  Therefore, at $k=0$ we can write
Eq.~(\ref{eq:cont2}) as
\beq
\label{eq:J(q,t)}
{\cal J}^z(k=0,q,t)=-{d\ov 2\sin(q/2)}\left[{H,\rho^{(2D)}(k=0,q,t)}\right].
\eeq
Evaluating the commutator with the help of Eqns.~(\ref{eq:multiCR}),
(\ref{eq:multirho,psi}), and (\ref{eq:tunnelop}), we find
\beq
\label{eq:Iz}
I^z(t)\equiv{\cal J}^z(k=0,q=0,t)
=-{i\lambda d\ov (2\pi a)^{\delta_{\lambda}}}\sum_j
\ix \left[{e^{i\alpha_{jq'}\phi_{q'}(x,t)}-\,\,{\rm h.c.}}\right].
\eeq
This expression will be used to calculate the $z$-axis conductivity of
the multilayer system.

\section{Marginal Tunneling Cases}
\label{sec:marginal}

We know there are only two maximally-chiral states with marginal
tunneling: 131 and the bosonic state 020.  In this section we begin
by showing that in the absence of disorder these states exhibit
insulating behavior in the vertical direction, \ie
$\sigma^{zz}(\omega)$ is identically zero for all frequencies at
$T=0$.  We will then find that adding disorder to the edge theory
increases the vertical conductivity, and therefore the surface behaves
like a ``chiral semi-metal''.  We conclude this section by considering
how the transport is modified by nearest-neighbor density-density
interactions.

\subsection{Clean Limit}
\label{ss:clean-marginal}

Our discussion of the system in the absence of disorder will proceed
in two steps.  First, we derive a sum rule which relates the frequency
integral of the conductivity to the expectation value of the tunneling
term in the Hamiltonian.  Next we prove that there exists a finite
range of values for the tunneling amplitude $\lambda$, and the
interaction strength $g$, for which the ground state in the presence
of tunneling and interactions is identical to the ground state in the
absence of tunneling and interactions.  We then combine these two
results to arrive at our conclusions about the zero temperature
conductivity.  We emphasize that this argument is non-perturbative in
$\lambda$ and $g$.  This is an important point since we do not have a
quadratic form of the Hamiltonian for the multilayer state with
marginal tunneling.

The conductivity sum rule is derived by considering the double
commutator of the Hamiltonian with the charge density operator.
Specifically, we calculate:
\beq
\label{eq:Hrhorho}
\left[{[{\cal H},\rho^{(2D)}(k=0,q,t)],\rho^{(2D)}(k=0,-q,t)}\right],
\eeq
where the Hamiltonian is given in Eq.~(\ref{eq:multiHfull}),
and from Eqns.~(\ref{eq:multirho,psi}), (\ref{eq:thetaq}), 
(\ref{eq:phiq}), (\ref{eq:FTrho}), and
$\rho_j^{(2D)}(x,t)=(1/d)\rho_j(x,t)$ we have
\beq
\label{eq:rhofromphiq}
\rho^{(2D)}(k=0,q,t)={1\ov 2\pi}\sqrt{N}\ix {1\ov\sqrt{K_q}}
\delx\phi_q(x,t).
\eeq
Using the commutation relations for the fields $\phi_q(x)$
(\ref{eq:phiqCR}) it is a straightforward exercise to evaluate the
commutator (\ref{eq:Hrhorho}) with the help of Eqns.~(\ref{eq:multiHfull}) and
(\ref{eq:rhofromphiq}).  One finds:
\beq
\label{eq:Hrhorho2}
\left[{[{\cal H},\rho^{(2D)}(k=0,q,t)],\rho^{(2D)}(k=0,-q,t)}\right]=
4\sin^2(q/2)\,{\cal H}_{\lambda},
\eeq
where ${\cal H}_{\lambda}$ is the tunneling part of the Hamiltonian.

We next relate the expectation value of this double commutator to the
integrated conductivity.  The continuity equation
(\ref{eq:continuity}) introduced in the previous section can be
written
\beq
\label{eq:continuity2}
\rho^{(2D)}(k=0,q,t)={2\ov d}\sin(q/2)\int {d\omega\ov 2\pi}
\int dt'\, {e^{-i\omega(t-t')}\ov \omega}{\cal J}^z(k=0,q,t).
\eeq
Suppressing the argument $k=0$, using $[{\cal H},\rho]=-i\del\rho/\del t$,
and taking the expectation value of Eq.~(\ref{eq:continuity2}) we find
\beq
\label{eq:sumrule1}
\left\la{\left[{[{\cal H},\rho^{(2D)}_j(q,t)],\rho^{(2D)}_j(-q,t)}\right]}\right\ra=
{4\ov d^2}\sin^2(q/2)\int {d\omega\ov 2\pi}\int dt\,e^{i\omega t}
\la[{\cal J}^z(q,0),{\cal J}^z(-q,t)]\ra.
\eeq
If we take the expectation value of Eq.~(\ref{eq:Hrhorho2}), and use it to
eliminate the l.h.s. of Eq.~(\ref{eq:sumrule1}) we find in the limit
$q\rightarrow 0$:
\beq
\label{eq:sumrule2}
\la {\cal H}_{\lambda}\ra={4\ov d^2}\int {d\omega\ov 2\pi}\int dt\,e^{i\omega t}
\la[I^z(0),I^z(t)]\ra,
\eeq
where we have used $I^z(t)={\cal J}^z(q=0,t)$.
The Kubo formula relates the commutator appearing in this expression
to the $z$-axis conductivity:
\beq
\label{eq:Kubo2}
\Re e\,\sigma^{zz}(\omega)={1\ov 2\omega}{1\ov NLd}\int dt\,
e^{i\omega t}\la [I^z(t),I^z(0)]\ra.
\eeq
Thus we arrive at the final form of the sum rule
\beq
\label{eq:sumrule3}
\int d\omega \, \Re e\, \sigma^{zz}(\omega)=-{\pi d\ov NL}\la {\cal H}_{\lambda}\ra.
\eeq
This is an exact relation valid at any temperature.  It applies to all
multilayer states, even in the presence of disorder, regardless of the
relevancy of the tunneling term.

Next we turn our attention to proving the stability of the ground
state in the presence of interlayer tunneling and interactions.  We
will describe in detail the case of the bosonic 020 state.  A proof
along the same lines can be constructed for the 131 state.  We begin
by writing the Hamiltonian of the 020 edge theory in terms of the
original radius $R=1$ Bose fields $u_j(x)$:
\beq
\label{eq:H020}
{\cal H}_{020}=\ix \sum_j\left[{
{v\ov 16\pi}\,:\!(\delx u_j)^2\!:\,+{g\ov 8\pi}\delx u_j\delx u_{j+1}
+{\lambda\ov (2\pi a)^2}\left({e^{i(u_j-u_{j+1})}+\,\,{\rm h.c.}}\right)
}\right].
\eeq
Instead of performing the discrete Fourier transformation along $z$
(\ref{eq:thetaq}) we simply rescale the fields, defining
$u_j(x)=\sqrt{2}\varphi_j(x)$, where $\varphi_j(x)$ are radius
$R=1/\sqrt{2}$ chiral bosons.  We can then define a set of $N$
$\widehat{su}(2)_1$ Kac-Moody (KM) currents:
\beq
\label{eq:multisu2}
J^z_j(x)={1\ov 2\pi\sqrt{2}}\delx\varphi_j(x),\qquad
J^{\pm}_j(x)=J^x_j \pm i J^y_j={1\ov 2\pi a}e^{\mp i\sqrt{2}\varphi_j(x)},
\eeq
in terms of which the Hamiltonian (\ref{eq:H020}) can be written
\beqarr
{\cal H}_{020}=\ix\left[{{\pi v\ov 6}
\,:\!\left[{{\bf J}_j(x)}\right]^2\!:\,+\pi g J^z_j(x)J^z_{j+1}(x)}\right.
\nonumber\\
\left.{{}+\lambda(J^-_j(x)J^+_{j+1}(x)+J^-_{j+1}(x)J^+_j(x))
}\right],
\label{eq:H020curr}
\eeqarr
where we have employed the identity
\beq
\label{eq:JztoJagain}
\ix \,:\!\left[{J^z(x)}\right]^2\!:\,
=\ix {1\ov 3}\,:\!\left[{{\bf J}(x)}\right]^2\!:.
\eeq
Consider the theory without interactions, $g=0$, and without tunneling
$\lambda=0$.  The zero-energy ground state of this Hamiltonian,
$|0\ra$, is formed by taking the tensor product of the highest-weight
state of the spin-0 irreducible representation of each
$\widehat{su}(2)_1$ algebra.  Thus, the ground state satisfies
\beq
\label{eq:gs_cond}
J_{j,n}^{a}|0\rangle=0,\,{\rm for}\quad \forall
\,\,j;\quad a=x,y,z;\quad n\geq 0,
\eeq
where the Fourier components of the currents are defined by
\beq
\label{eq:fourierJ}
J^a_{j,n}\equiv \ix J^a_j(x)e^{-2\pi i n x/L},\quad n\in\Zint,
\eeq
and obey the algebra
\beq
\label{eq:KMA}
[J^a_{i,n}, J^b_{j,m}]={n\over 2}\delta_{ij}\delta^{ab}\delta_{n+m,0}
+i\delta_{ij}\epsilon^{abc}J^c_{j,n+m}.
\eeq
Using Eqns.~(\ref{eq:H020curr}) and (\ref{eq:fourierJ}) we can write the
Hamiltonian as
\beq
\label{eq:H020curr2}
{\cal H}_{020}=\left[{{\pi v\ov 6L}
\,:\!J^a_{j,n}J^a_{j,-n}\!:\,+{\pi g\ov L} J^z_{j,n}J^z_{j+1,-n}
+{\lambda\ov L}(J^-_{j,n}J^+_{j+1,-n}+J^-_{j+1,n}J^+_{j,-n})
}\right].
\eeq
For later reference we also record the expression for the current
operator in terms of the KM generators
\beq
\label{eq:Izcurr}
I^z=-{i\lambda d\ov L}\left[{
J^-_{j,n}J^+_{j+1,-n}-J^-_{j+1,n}J^+_{j,-n}
}\right].
\eeq

From Eqns.~(\ref{eq:gs_cond}) and (\ref{eq:H020curr2}) we see that not
only is $|0\ra$ the zero-energy ground state of the Hamiltonian with
$g=\lambda=0$, it is a zero-energy eigenstate of the Hamiltonian for
all $g$ and $\lambda$:
\beq
\label{eq:H020ann0}
{\cal H}_{020}|0\ra = 0.
\eeq

Of course, this does not mean that
$|0\ra$ is the ground state of the Hamiltonian with non-zero $g$ and
$\lambda$.  To establish this we now demonstrate that for a range of
$g$ and $\lambda$ the Hamiltonian is positive semi-definite.  For
$\lambda>0$, $g>0$, we rewrite Eq.~(\ref{eq:H020curr}) as 
\beqarr
\label{eq:H020curr3}
{\cal H}_{020}&=&\ix\biggl\{{\pi g\ov 2}
:\!\left({J}^z_j+{J}^z_{j+1}\right)^2\!:\,+ {\lambda\ov
  2}\left[:\!\left({J}^-_j+{J}^-_{j+1}\right)
    \left({J}^+_j+{J}^+_{j+1}\right)\!: +(J^+\leftrightarrow J^-)
  \right]\nonumber\\
& \relax& + \left({\pi v\ov 2}-\pi g\right)
\,:\!\left[{J^z_j(x)}\right]^2\!: -
\lambda\left[:\!{J}^-_j(x){J}^+_j(x)\!:
    +:\!{J}^+_j(x){J}^-_j(x)\!:\right] \biggr\}.  
\eeqarr 
The two terms in the first line are clearly non-negatively defined,
while the terms in the second line can be shown to be non-negatively
defined for $v>2g+8\lambda/\pi$ using Eq.~(\ref{eq:JztoJagain}) and
the basic identity
\beq
\label{eq:spins}
\left({{\bf J}_j}\right)^2=\left({J^z_j}\right)^2+
{1\ov 2}(J^+_jJ^-_j+J^-_jJ^+_j).
\eeq
An analogous construction can be done for $\lambda<0$ or $g<0$, and we
conclude that the Hamiltonian~(\ref{eq:H020curr}) is positive
semi-definite provided
\beq
\label{eq:restrictglambda}
2|g|+{8}|\lambda|/\pi<{v}.  
\eeq 
This is also the sufficient condition for the stability of the
original ground state, as follows from our previous result that the
Hamiltonian annihilates the original ground state $|0\rangle$.  

Note that the normal ordering does not present a complication because
it amounts to subtracting off the unit operator times the expectation
value of the Hamiltonian in the state $|0\ra$, which is zero.
Therefore, since the Hamiltonian is positive semi-definite and we know
the state $|0\ra$ is a zero-energy eigenstate, it follows that $|0\ra$
is the ground state of the Hamiltonian provided
Eq.~(\ref{eq:restrictglambda}) is satisfied.

We can now combine the sum rule with our knowledge of the exact ground
state to say something about the conductivity.  At zero temperature
the expectation value on the r.h.s. of the sum rule
(\ref{eq:sumrule3}) is a ground state expectation value.  Since we
know the ground state $|0\ra$ is unchanged by tunneling and
interactions and $\la 0|{\cal H}_1|0 \ra=0$ we can
conclude
\beq
\label{eq:intsigma0}
\int d\omega \, \Re e\,\sigma^{zz}(\omega)=0.
\eeq
The real part of $\sigma^{zz}(\omega)$ is dissipative and hence it
cannot be negative.  Therefore, from Eq.~(\ref{eq:intsigma0}) we have
\beq
\label{eq:sigma0}
\sigma^{zz}(\omega)=0.  
\eeq 
We reiterate that this is an exact statement for the clean 020
multilayer at $T=0$ provided $g$ and $\lambda$ satisfy
Eq.~(\ref{eq:restrictglambda}).  As we mentioned above, a similar
result can be shown to hold for the 131 state, the other
maximally-chiral state with marginal tunneling.  The condition for the
stability of the ground state of the 131 edge theory is also given by
Eq.~(\ref{eq:restrictglambda}).  In the next section we will explore
what happens when we add disorder to these systems.

\subsection{Disordered Case}
\label{ss:dirty-marginal}

In the previous section we found that the clean multilayer with
marginal tunneling exhibits insulating behavior.  In this section we
study the marginal tunneling case in the presence of disorder.  We
perform a perturbative calculation of the vertical conductivity via
the Kubo formula, finding that disorder increases the conductivity.

In Section~\ref{sec:edge} we discussed the general edge theory of
clean multilayer systems.  Our first task is to add disorder to this
model.  We will then use the Kubo formula to find the $z$-axis
conductivity.

The disorder we consider is a random scalar potential $V_j(x)$ in each
layer $j$, which couples to the Bose fields via the charge density
\beq
\label{eq:rhoV}
{\cal H}_V=\ix V_j(x)\rho_j(x)=\ix {1\ov 2\pi}V_j(x)K_{jk}\delx u_k(x).
\eeq
We assume $V_j(x)$ is a Gaussian random variable, uncorrelated
between different layers, but for now we will not specify how it is
correlated within the same layer.

Using the Fourier transformation (\ref{eq:thetaq}) and rescaling
(\ref{eq:phiq}) of the Bose fields we can write Eq.~(\ref{eq:rhoV}) as
\beq
\label{eq:rhoV2}
{\cal H}_V=\ix {1\ov 2\pi}{1\ov \sqrt{K_q}}V_q\delx\phi_{-q}(x), 
\eeq
where
\beq
\label{eq:VQ}
V_q(x)={1\ov \sqrt{N}}\sum_j e^{-iqj}V_j(x).
\eeq
Recalling the form of the clean multilayer edge theory from
Section~\ref{sec:edge}, our full Hamiltonian is
\beqarr
{\cal H}&=&\ix \left[{{1\over{4\pi}} v_q 
\,:\!\delx\phi_q\delx\phi_{-q}\!:\,+\sum_j{\lambda\ov (2\pi a)^{2}}
\left({e^{i\alpha_{jq}\phi_q(x)}+\,\,{\rm h.c.}}\right)}\right.
\nonumber\\
&&\left.{{}+{1\ov 2\pi}{1\ov \sqrt{K_q}}V_q\delx\phi_{-q}(x)
}\right].\label{eq:multiHwdirt}
\eeqarr
We move the disorder term into the phase of the tunneling
amplitude by shifting the bosons:
\beq
\label{eq:shiftphiq}
\phi_q\mapsto \phi_q+ {1\ov v_q\sqrt{K_q}}\int^x dy\, V_q(y).
\eeq
If we define the phases
\beq
\label{eq:gammaj}
\gamma_j(x)\equiv-{1\ov\sqrt{N}}\sum_q e^{iqj}(1-e^{iq}){1\ov v_q}
\int^x dy\, V_q(y),
\eeq
then with Eq.~(\ref{eq:shiftphiq}) the Hamiltonian
(\ref{eq:multiHwdirt}) is
\beq
\label{eq:multiHwdirt2}
{\cal H}=\ix \left[{{1\ov 4\pi}v_q 
\,:\!\delx\phi_q\delx\phi_{-q}\!:\,+{\lambda\ov (2\pi a)^{2}}
\sum_j\left({e^{i\gamma_j(x)}e^{i\alpha_{jq}\phi_q(x)}
+\,\,{\rm h.c.}}\right)}\right],
\eeq
and the current operator (\ref{eq:Iz}) becomes
\beq
\label{eq:dirtyIz}
I^z=-{i\lambda d\ov (2\pi a)^{2}}\sum_j
\ix \left[{e^{i\gamma_j(x)}e^{i\alpha_{jq}\phi_{q}(x)}
-\,\,{\rm h.c.}}\right].
\eeq

We are now in a position to calculate the $z$-axis conductivity.  We
will first evaluate the Matsubara two-point function of the current
operator (\ref{eq:dirtyIz}).  Since Eq.~(\ref{eq:multiHwdirt2}) is a
theory of $N$ interacting chiral bosons, the best we can do is a
perturbative calculation in the tunneling amplitude $\lambda$.
Although we will work perturbatively in $\lambda$, we will be treating
the disorder exactly.  We thus expect the ${\cal O}(\lambda^2)$ result
to be reliable, since the disorder essentially randomizes the phase of
the electrons between tunneling events.  More formally, if the
tunneling amplitude is a coordinate-dependent, delta-correlated random
variable, then the RG flow (\ref{eq:multiflow}) is modified to
\beq
\label{eq:randomflow}
{d\Delta \ov d\ell}=(3-2\delta_{\lambda})\Delta,
\eeq
where $\Delta$ is the variance of the tunneling.
Thus for $\delta_{\lambda}=2$ we would conclude that the
tunneling is irrelevant.  The combination $\lambda e^{i\gamma_j(x)}$
that appears in the Hamiltonian (\ref{eq:multiHwdirt2}) is in general
not delta-correlated, as we will discuss below, but since the case of
$x$-independent tunneling is marginal we expect any departure from
uniformity makes the tunneling irrelevant and therefore justifies a
perturbative expansion in $\lambda$.

Using the expression for the current operator (\ref{eq:dirtyIz}) 
we evaluate the Matsubara two-point function
\beqarr
{\mat C}(\tau)&=&-\la T_{\tau}I^z(\tau)I^z(0)\ra\nonumber\\
&=&{d^2\lambda^2\ov L^4}\int dx\,dx'\,\sum_{j,k}
\left\la{T_{\tau}\left({
e^{i\gamma_j(x)}\,:\!e^{i\alpha_{jq}\phi_q(\tau,x)}\!:\,
-\,e^{-i\gamma_j(x)}\,:\!e^{-i\alpha_{jq}\phi_q(\tau,x)}\!:
}\right)}\right.\nonumber\\
&&\left.{{}\times\left({
e^{i\gamma_k(x')}\,:\!e^{i\alpha_{kq}\phi_q(0,x')}\!:\,
-\,e^{-i\gamma_k(x')}\,:\!e^{-i\alpha_{kq}\phi_q(0,x')}\!:
}\right)}\right\ra,\label{eq:margC}
\eeqarr
where we have normal-ordered the vertex operators.  
Note that because the current
operator (\ref{eq:dirtyIz}) has a factor of $\lambda$ out front, there
is an explicit factor of $\lambda^2$ in Eq.~(\ref{eq:margC}), and
therefore to find this function to order $\lambda^2$ we can evaluate
the expectation value using the $\lambda=0$ Hamiltonian.  One
finds that only the cross terms in the product of the current
operators give non-vanishing contributions, and only when $j=k$.  The
terms have the form
\beqarr
\lefteqn{-\left\la{T_{\tau}\,:\!e^{i\alpha_{jq}\phi_q(\tau,x)}\!:\,
:\!e^{-i\alpha_{jq}\phi_q(0,x')}\!:}\right\ra}\nonumber\\
&&=\prod_q{L^{\alpha^2_q}\ov \left[{(2\beta i v_q)
\sinh\left({{\pi\ov\beta v_q}(x-x'+iv_q\tau+ia\,\sgn\,\tau)
}\right)}\right]^{\alpha_q^2}},\label{eq:IzIz}
\eeqarr
where $\alpha_q^2\equiv \alpha_{jq}\alpha_{j-q}$, with no sum on $j$.
This is a very complicated
function; a great simplification occurs if the eigenmode velocities,
$v_q$, are identical for all $q$.  From the expression for $v_q$ 
(\ref{eq:v_q}) we see that this occurs when $g=nv/m$.  For the
two multilayer states with marginal tunneling, 020 and 131, this
condition gives $g=0$ and $g=v/3$, respectively.  We will refer to the
case where $v_q$ is independent of $q$ as the ``solvable point.''  For
020 it corresponds to no density-density couplings between neighboring
layers while for 131 it occurs at a non-zero value of the
density-density interaction strength.  Note that for more general
forms of the matrices $K$ and $U$, \ie not tridiagonal, the solvable
point corresponds to $U\propto K$.

Working at the solvable point, writing $v_q\equiv v_0$, and using our
knowledge of the scaling dimension of the tunneling operator 
(\ref{eq:deltalambda}),
\beq
\label{eq:sumalphas2}
\sum_q \alpha_q^2=\sum_q \alpha_{jq}\alpha_{j-q}=2\delta_{\lambda}=4,
\eeq
we find upon disorder averaging Eq.~(\ref{eq:margC})
\beq
\label{eq:margCavg}
\ol{\mat C}(\tau)=-{2d^2\lambda^2\ov (2\beta v_0)^4}\sum_j\int dx\,dx'\,
{W_j(x,x')\ov \left[{\sinh\left({{\pi\ov\beta v_0}
(x-x'+iv_0\tau+ia\,\sgn\,\tau)}\right)}\right]^4},
\eeq
where we have defined the function
\beq
\label{eq:W(x)}
W_j(x,x')\equiv \ol{e^{i[\gamma_j(x)-\gamma_j(x')]}}.
\eeq
If we take the correlation function of the scalar potential to be
\beq
\label{eq:Vcorr}
\ol{V_j(x)V_k(x')}=\delta_{jk}Z(x-x'),
\eeq
then using the definition of the phase $\gamma_j(x)$ (\ref{eq:gammaj}) we
find that the function $W_j(x,x')$ can be written
\beq
\label{eq:W(x)2}
W_j(x,x')=\exp\left[{-{1\ov v_0^2}\int_{x}^{x'}dy_1\,
\int_{x}^{x'}dy_2\, Z(y_1-y_2)}\right],
\eeq
Note that $W_j(x,x')\equiv W(x-x')$ is independent of $j$ and depends
only on the magnitude of the relative coordinate $|x-x'|$.  Therefore
the sum over $j$ in Eq.~(\ref{eq:margCavg}) just gives a factor of the
number of layers, $N$, and we can replace the integration over $x$ and
$x'$ by an integration over the average coordinate, which gives a
factor of $L$, and an integration over the relative coordinate
$y=x-x'$.  If we define the Fourier transform of the function $W(x)$:
\beq
\label{eq:W(k)}
\Wt(k)\equiv\int dx\, e^{-ikx}W(x),
\eeq
then we can write Eq.~(\ref{eq:margCavg}) as
\beq
\label{eq:margCavg2}
\ol{\mat C}(\tau)=-{2d^2NL\lambda^2\ov (2\beta v_0)^4}\int {dk\ov 2\pi}
\Wt(k)\int dy\,{e^{iky}\ov \left[{\sinh\left({{\pi\ov\beta v_0}
(y+iv_0\tau+ia\,\sgn\,\tau)}\right)}\right]^4}
\eeq
The factor of $L$ in this expression will cancel against the factor of
$1/L$ in the Kubo formula (\ref{eq:Kubo2}), and in anticipation of this
we have extended the range of the $y$ integral to include the entire
real axis since we are interested in the result in the limit
where the system size $L\rightarrow \8$.  The integration over $y$ can
now be performed by the method of residues.  The integrand has
fourth-order poles on the imaginary $y$-axis at the points
$y_n=i[v_0(\beta n-\tau)-a\,\sgn\,\tau]$ for integer $n$, and is
exponentially small in the upper (lower) half plane for $k>0$
($k<0$).  One finds
\beq
\label{eq:intsinh}
\int dy\,{e^{iky}\ov \left[{\sinh\left({{\pi\ov\beta v_0}
(y+iv_0\tau+ia\,\sgn\,\tau)}\right)}\right]^4}
={\pi\ov 3}\left[{4\left({\beta v_0k\ov\pi}\right)+
\left({\beta v_0k\ov\pi}\right)^3}\right]
{e^{v_0k\tau}\ov e^{\beta v_0k}-1}.
\eeq
We now Fourier transform Eq.~(\ref{eq:margCavg2}) with respect to the
imaginary time $\tau$,    
\beqarr
\ol{\mat C}(\omega_n)&=&\int_0^{\beta}d\tau\,e^{i\omega_n\tau}
\ol{\mat C}(\tau)\nonumber\\
&=& -{d^2\lambda^2NL\ov 48\pi^4v_0^4}
\int dp\,\Wt(p/v_0){p^3+4\pi^2p/\beta^2\ov i\omega_n+p},
\label{eq:margC(omega)}
\eeqarr
where we have changed the integration variable to $p\equiv v_0k$.
Analytically continuing ($i\omega_n\mapsto \omega+i0^+$) we find
\beqarr
\ol{\cal C}(\omega)&=&-i\int_0^{\8}dt\,e^{i\omega t}
\ol{\la[I^z(t),I^z(0)]\ra}\nonumber\\
&=& -{d^2\lambda^2NL\ov 48\pi^4v_0^4}
\int dp\,\Wt(p/v_0){p^3+4\pi^2p/\beta^2\ov \omega+p+i0^+}.
\label{eq:margC(omega)2}
\eeqarr
Using this result in the Kubo formula (\ref{eq:Kubo2}) with the help
of the identity
\beq
\label{eq:PViden}
{1\ov x\pm i0^+}= {\cal P}{1\ov x} \mp i\pi\delta(x),
\eeq
and the observation that $\Wt(k)$ is real since $W(x)$ is an even
function of $x$, we arrive finally at
\beq
\label{eq:sigmazzmarg}
\Re e \,\ol{\sigma^{zz}}(\omega)={\lambda^2 d\ov 48\pi^3 v_0^4}
(\omega^2+4\pi^2T^2)\Wt(\omega/v_0).
\eeq
This is our central result.  It is the
${\cal O}(\lambda^2)$ term in the vertical conductivity at the solvable point 
for both the 020 and 131 states at a finite temperature $T$,
including potential disorder which enters through the function $\Wt(k)$.
The imaginary part of the conductivity can of course be found from
Eq.~(\ref{eq:sigmazzmarg}) by the usual dispersion relations\cite{mahan}.

Let us first consider the limit of a clean system, $V_j(x)\equiv 0$.
From Eqns.~(\ref{eq:Vcorr}) and (\ref{eq:W(x)2}), we see that in this
case $W(x)=1$ and therefore $\Wt(k)=2\pi\delta(k)$.  Hence from
Eq.~(\ref{eq:sigmazzmarg}) we find
\beq
\label{eq:cleansigmazz}
\Re e\, \sigma^{zz}(\omega)={\lambda^2 d\ov 6v_0^3}T^2\delta(\omega).
\eeq 
This is consistent with our result in Section~\ref{ss:clean-marginal}
that the $z$-axis conductivity in the clean system vanishes at $T=0$.
An interesting aspect of this result is that it is proportional to
$\delta(\omega)$, \ie the conductivity vanishes at all $\omega\neq 0$,
and the DC conductivity is infinite.  One question to ask is whether
these features are a consequence of the fact that
Eq.~(\ref{eq:cleansigmazz}) is only the first non-vanishing term in a
perturbative expansion in $\lambda$.

Although we cannot definitively answer the question about the
finiteness of the DC conductivity beyond ${\cal O}(\lambda^2)$, we can
prove that at higher orders in $\lambda$ the conductivity cannot
vanish at all $\omega\neq 0$.  First we note that in the clean system
the current operator commutes with the Hamiltonian in the absence of
tunneling: $[I^z,{\cal H}_0]=0$.  For example, for the 020 state this
can be seen from Eqns.~(\ref{eq:H020curr2}) and (\ref{eq:Izcurr}),
using the commutation relations for the Kac-Moody currents given in
Eq.~(\ref{eq:KMA}).  Since the current operator commutes with
${\cal H}_0$, the expectation value of the current-current commutator
evaluated using the $\lambda=0$ Hamiltonian vanishes, and therefore by
the Kubo formula (\ref{eq:Kubo2}) the vertical conductivity must
vanish for all non-zero frequencies.  This explains why the
conductivity is zero except at $\omega=0$ in
Eq.~(\ref{eq:cleansigmazz}).  By the same reasoning we see that if the
current operator commutes with the full Hamiltonian, $[I^z,{\cal
H}]=0$, then $\sigma^{zz}(\omega)$ would be zero for all $\omega\neq
0$ to {\em all} orders in $\lambda$.  The converse of this is also true, \ie
if $\sigma^{zz}(\omega)$ is zero for all $\omega\neq0$ then
$[I^z,{\cal H}]=0$.  This is proven in Appendix~\ref{sec:converse}.
For the multilayer edge theory with marginal tunneling: $[I^z,{\cal
H}]\neq 0$.  For the 020 case this can be established using
Eqns.~(\ref{eq:H020curr2}) and (\ref{eq:Izcurr}).  Therefore, since
the current operator does not commute with the full Hamiltonian, we
can conclude that $\sigma^{zz}(\omega)$ cannot be zero for all $\omega
\neq 0$ for the clean 020 or 131 multilayers.  Hence, we would expect
that the $\delta(\omega)$ factor in Eq.~(\ref{eq:cleansigmazz}) would
be broadened by the higher-order corrections in $\lambda$.  Recalling
that $\lambda$ is dimensionless and the conductivity should not depend
on the sign of $\lambda$, one plausible scenario is:
\beq
\label{eq:broadendelta}
\Re e\,\sigma^{zz}(\omega)={\lambda^2 d\ov 6v_0^3}T\delta(\omega/T)
\mapsto \,\,\,\,\sim{\lambda^2 d\ov 6v_0^3}T{\lambda^2\ov (\omega/T)^2+{\cal
O}(\lambda^4)}.
\eeq

This conjectured form has several interesting features.  First, we
find that the DC conductivity goes to zero with temperature as $T$.
We expect this to be a generic feature of the exact result in
$\lambda$ for the clean system, provided the DC conductivity is
finite.  This differs from the disordered system
(\ref{eq:sigmazzmarg}) where we find the DC conductivity vanishes as
$T^2$.  In addition, note that the $\omega\rightarrow 0$ limit of
Eq.~(\ref{eq:broadendelta}) is independent of the tunneling amplitude
$\lambda$.

We now return to the result for $\ol{\sigma^{zz}}(\omega)$ in the
presence of disorder (\ref{eq:sigmazzmarg}).  If the disorder is
delta-correlated, the function $Z(x)$ defined in
Eq.~(\ref{eq:Vcorr}) is
\beq
\label{eq:deltaVq}
Z(x)=\Delta\delta(x),
\eeq
where $\Delta$ is the variance of the disorder.  For this case
$W(x)=\exp(-\Delta|x|/v_0^2)$, which has a Fourier transform 
$\Wt(k)=(2\Delta/v_0^2)/(k^2+(\Delta/v_0^2)^2)$, and hence
(\ref{eq:sigmazzmarg}) yields
\beq
\label{eq:sigmazzmarg2}
\Re e\, \ol{\sigma^{zz}}(\omega)={\lambda^2 d\Delta\ov 24\pi^3 v_0^4}
{\omega^2+4\pi^2T^2\ov \omega^2+(\Delta/v_0)^2}.
\eeq
Note that at zero temperature the conductivity vanishes in the DC
limit.  This conclusion is independent of the details of the disorder
and holds as long as $\Wt(0)$ is finite.  We also see that the
conductivity approaches a constant as $\omega\rightarrow \8$.  This
second feature is a consequence of taking the disorder to be
delta-correlated.  We see that $\ol{\sigma^{zz}}(\omega)$ goes to a
constant at large $\omega$ because of the slow fall off of $\Wt(k)$,
which is in turn caused by the non-analyticity of $W(x)$ at $x=0$.  We
shall now investigate how Eq.~(\ref{eq:sigmazzmarg2}) is modified if
the disorder has a finite correlation length.

As a first step, we determine the asymptotic form of the
disorder-averaged phase, $W(x)$, at small and large $x$ when the
scalar potential disorder has a finite correlation length.  We return
to the expression for the correlation function of the random potential
(\ref{eq:Vcorr}) and write
\beq
\label{eq:paramZ}
Z(x)\equiv \Delta^2A(x/\ell_0),
\eeq
where $\Delta$ is a parameter with the dimensions of energy that sets
the strength of the disorder, $\ell_0$ is the correlation length of
the disorder, and $A(z)$ is a dimensionless function which we take to
have the properties: $A(0)=1$, $A(-z)=A(z)$, and 
\beq
\label{eq:intA}
\int_0^{\8}dz\, A(z)=1.
\eeq
Now consider the disorder-averaged phase, $W(x)$, given in
Eq.~(\ref{eq:W(x)2}).  For $|x|\ll \ell_0$, \ie for distances
small compared with the correlation length of the random potential, we
can replace $Z(y_1-y_2)$ by its value at $y_1-y_2=0$ and obtain:
\beq
\label{eq:Watsmallx}
W(\ell_0z)\approx e^{-(\Delta\ell_0z/v_0)^2}\approx
1-\left({\Delta\ell_0z\ov v_0}\right)^2,\quad
{\rm for}\quad |z|\ll 1.
\eeq
In the opposite limit $|x|\gg \ell_0$ we change integration variables 
in Eq.~(\ref{eq:W(x)2}) to $y_c=(y_1+y_2)/2$ and $y_r=y_1-y_2$:
\beq
\label{eq:Wchanged}
W(x)=\exp\left[{
-{1\ov v_0^2}\left({\int_{-|x|}^0dy_r\,(|x|+y_r)Z(y_r)+\int_0^{|x|}dy_r\,
(|x|-y_r)Z(y_r)}\right)}\right].
\eeq
Since $|x|/\ell_0 \gg 1$ we can extend the integration over $y_r$ to
infinity with small error.  Using the symmetry of $A$ and defining
\beq
\label{eq:ell12}
\zeta=\int_0^{\8}dz\,zA(z),
\eeq
we then find
\beq
\label{eq:Watlargex}
W(\ell_0z)\approx \exp
\left[{-2\left({\Delta\ell_0\ov v_0}\right)^2(|z|-\zeta)}\right],
\quad {\rm for}\quad |z|\gg 1.
\eeq
We see that at short distances $W(x)$ is parabolic
(\ref{eq:Watsmallx}) while at long distances it decays exponentially
(\ref{eq:Watlargex}).  The presence of a finite
correlation length for the random potential does not change the long
distance behavior of $W(x)$, but it does ``round out'' the
non-analyticity at $x=0$ present in the case of delta-correlated
disorder.  Generally, this is sufficient to make the function rapidly
decaying as $\omega \rightarrow \8$.

Having found the generic behavior of the function $W(x)$, we now
choose a specific form:
\beq
\label{eq:modelW}
W(x)={1\ov\cosh(\xi x)},
\eeq
characterized by the parameter $\xi$ which has the dimensions of an
energy.  This choice of $W(x)$ has several appealing features.  First,
it has the correct asymptotic forms at large and small values of $x$.
Indeed, if we match the asymptotics of Eq.~(\ref{eq:modelW}) to
Eqns.~(\ref{eq:Watsmallx}) and (\ref{eq:Watlargex}) we find
\beq
\label{eq:relateparam}
{\Delta\ov v_0}=\xi,\quad \ell_0={1\ov 2\xi}.
\eeq
We see that for the choice of $W(x)$ in Eq.~(\ref{eq:modelW}), the
parameter $\xi$ is proportional to the strength of the disorder and
inversely proportional to the correlation length.  Hence this form
allows us to explore the effect of a finite correlation length in the
region $\Delta\ell_0\sim 1$.

\begin{figure}[htbp]
  \begin{center}
    \leavevmode
    \epsfxsize=0.6\columnwidth
    \epsfbox{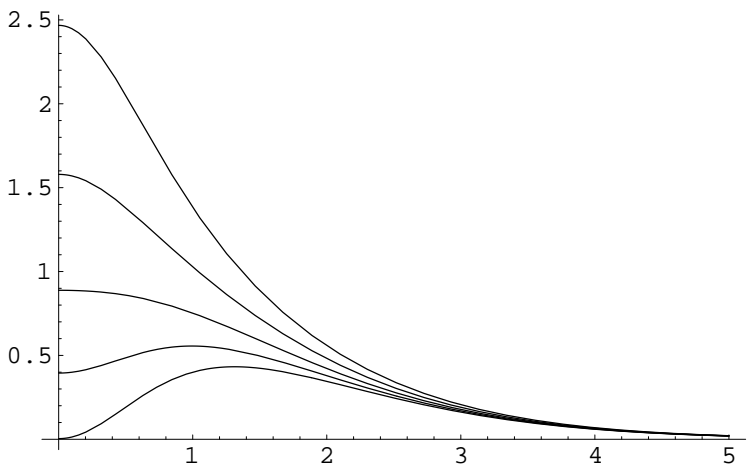}\vskip1mm
    \caption{
    The real part of the vertical conductivity for the 020 or 131
    multilayer with disorder (\protect\ref{eq:sigmazzwell0}).  The horizontal
    axis is frequency measured in units of the parameter
    $v_0\xi$.  The temperatures of the curves, starting with the
    uppermost, are: $T/v_0\xi=0.25,0.2,0.15,0.1,0.01$.}
    \label{fig:sigmazz}
  \end{center}
\end{figure}

A second advantage of Eq.~(\ref{eq:modelW}) is that its
Fourier transform can be evaluated in closed form.  A straightforward
computation gives
\beq
\label{eq:modelWt}
\Wt(k)=\int dx\, {e^{-ikx}\ov \cosh (\xi
x)}={\pi/\xi\ov\cosh\left({\pi k/2\xi}\right)}.
\eeq
Using this result in the general form above for the conductivity
(\ref{eq:sigmazzmarg}) we find
\beq
\label{eq:sigmazzwell0}
\Re e \,\ol{\sigma^{zz}}(\omega)={\lambda^2 d\ov 48\pi^2 v_0^4}
{\omega^2+4\pi^2T^2\ov \xi\cosh\left({\pi \omega/2\xi v_0}\right)}.
\eeq
Comparing Eq.~(\ref{eq:sigmazzwell0}) with the expression for the
conductivity in the case of delta-correlated disorder
(\ref{eq:sigmazzmarg2}) we see that the finite correlation length
gives a conductivity which exponentially decays to zero as
$\omega\rightarrow \8$ rather than approaching a nonzero value.  This
result for $\ol{\sigma^{zz}}$ (\ref{eq:sigmazzwell0}) is shown in
Fig.~\ref{fig:sigmazz} at various temperatures.  For large
temperatures the conductivity is peaked around $\omega=0$, while at
smaller temperatures the maximum occurs at a finite frequency.

\subsection{Effect of Interactions}
\label{ss:interactions}

In the preceding section we calculated the order $\lambda^2$ term
in the $z$-axis conductivity of the multilayer states with marginal
tunneling, 020 and 131, at the solvable point where the eigenmode
velocities $v_q$ are independent of $q$.  In this section we discuss
how the zero-temperature result is modified away from the solvable
point.

Since we will only consider $T=0$ in this section we shall work
directly in real time $t$.  The real-time zero-temperature analog of
the expression  for the Matsubara two-point function of the
current operator (\ref{eq:margCavg}) is
\beq
\label{eq:realC}
\ol{\cal C}(t)=-{2id^2\lambda^2NL\ov (2\pi)^4}\int dy\,
W(y)\prod_q {1\ov (y-v_qt+ia\,\sgn\,t)^{\alpha_q^2}},
\eeq
where from the definitions (\ref{eq:gammaj}) and (\ref{eq:W(x)}) we
see that the expression for $W(x)$ in Eq.~(\ref{eq:W(x)2}) is changed to
\beq
\label{eq:intW}
W(x)=\exp\left[{-\left({{2\ov N}\sum_q{\sin^2(q/2)\ov v_q^2}}\right)
\int^{x}_{0}dy_1\,\int^{x}_{0}dy_2\, Z(y_1-y_2)}\right].
\eeq
The corresponding retarded correlation function is
\beq
\label{eq:retardC}
\ol{{\cal C}^R}(t)=-{2id^2\lambda^2NL\ov (2\pi)^4}\theta(t)\int dy\,
W(y)\left[{\prod_q {1\ov (y-v_qt+ia)^{\alpha_q^2}}
-\prod_q {1\ov (y-v_qt-ia)^{\alpha_q^2}}
}\right].
\eeq

We would now like to perform the $y$-integration.  From
Eq.~(\ref{eq:alphas}) we see that the exponents appearing in
Eq.~(\ref{eq:retardC}) are
\beq
\label{eq:expon}
\alpha_q^2={2\ov N}(3-\cos q-2\cos^2 q),
\eeq
and therefore, as a function of complex $y$, the first integrand in
Eq.~(\ref{eq:retardC}) has $N$ branch point singularities at the
points $v_qt-ia$ and similarly for the second integrand at
$v_qt+ia$.  This is a complicated singularity structure, but
observe that in the first term these branch points are all below the
real axis, while in the second term they are all above the real axis.
Hence, since $\sum_q \alpha_q^2$ is an integer,
we can certainly choose branch cuts that lie entirely on one
side of the real axis for each term.  When we considered the solvable
point in the previous section we were able to evaluate the
conductivity for any disorder $W(y)$.  We shall not attempt the same
feat away from the solvable point.  The evaluation of the $y$-integral
in Eq.~(\ref{eq:retardC}) is greatly simplified if $W(y)$ is a
meromorphic function of $y$.  One such function was discussed in the
previous section (\ref{eq:modelW}), and for the remainder of this
section we specialize to this form.

Using $W(y)=1/\cosh(\xi y)$ in Eq.~(\ref{eq:retardC}), we can close
the first term in the upper-half plane and the second term in the
lower-half plane, avoiding all the complicated branch point
singularities, and picking up the residues of the poles of $W(y)$.  We
find
\beqarr
\ol{{\cal C}^R}(t)&=&-i\theta(t)\la \ol{[I^z(t),I^z(0)]}\ra\nonumber\\
&=&-{2id^2\lambda^2NL\ov (2\pi)^3}\theta(t)\xi^3
\sum_{r=-\8}^{\8}(-1)^r\prod_q {1\ov [v_q\xi t-i\pi(r+1/2)]^{\alpha_q^2}}.
\label{eq:retardC2}
\eeqarr
Therefore, from the Kubo formula (\ref{eq:Kubo2}) we arrive at
\beq
\label{eq:sigmazzwint}
\Re e\, \ol{\sigma^{zz}}(\omega)= {\lambda^2 d\ov
(2\pi)^3}{\xi^3\ov\omega}
\sum_{r=-\8}^{\8}(-1)^r\int dt\,e^{i\omega t}
\prod_q {1\ov [v_q\xi t-i\pi(r+1/2)]^{\alpha_q^2}}.
\eeq
This is certainly far from a closed form result, but it is a
convenient form for extracting the high and low frequency asymptotics
of the conductivity.  At large (positive) $\omega$ the asymptotic form
is determined by the closest singularity to the real axis in the
upper half of the complex $t$ plane.  From Eq.~(\ref{eq:sigmazzwint})
this clearly occurs a finite distance away from the real-$t$ axis 
and we find
\beq
\label{eq:largeomegacond}
\Re e\, \ol{\sigma^{zz}}(\omega)
\stackrel{\omega\rightarrow\8}{\longrightarrow}
\exp\left[{-{\pi\omega\ov 2\xi v_{\rm max}}}\right],
\eeq
where $v_{\rm max}$ is the maximum of $v_q$ over $q$.  For the 020
state we see from the expression for $v_q$ (\ref{eq:v_q}) that at the
solvable point ($g=0$) $v_0=v/2$ while away from the solvable point 
($g>0$) $v_{\rm max}=v/2+g$.  Comparing Eq.~(\ref{eq:largeomegacond})
with the asymptotic form of Eq.~(\ref{eq:sigmazzwell0}):
\beq
\label{eq:largeomegacond2}
\Re e\, \ol{\sigma^{zz}}(\omega)
\stackrel{\omega\rightarrow\8}{\longrightarrow}
\exp\left[{-{\pi\omega\ov 2\xi v_0}}\right],
\eeq 
we find that one effect of interactions at $T=0$ is to soften the
exponential decay of the conductivity at large frequencies.

Returning to Eq.~(\ref{eq:sigmazzwint}), we now consider the behavior
at small frequencies.  If we assume $\omega>0$ we can close the $t$
integral in the upper-half plane because the integrand is
exponentially small there.  Hence all the terms with $r<0$ give no
contribution.  For each $r\geq 0$ there remains $N$ branch point
singularities in the upper-half plane enclosed by the contour.  Since
we are interested in small $\omega$ we stretch the
integration contour into a large circle so that everywhere along
the contour $|t|$ is very large.  We can then expand the integrand in
powers of $1/t$ and integrate term-by-term:
\beqarr
\lefteqn{\int dt\,e^{i\omega t}\prod_q {1\ov [v_q\xi
t-i\pi(r+1/2)]^{\alpha_q^2}}}\nonumber\\
&&=\oint dt\,e^{i\omega t}\left({\prod_q {1\ov
v_q^{\alpha_q^2}}}\right)
{1\ov (\xi t)^4}\left[{1+{i\pi(r+1/2)\ov\xi t}\sum_q{\alpha_q^2\ov v_q}
+{\cal O}\left({1\ov t^2}\right)
}\right]\nonumber\\
&&={\pi\ov 3}\left({\prod_q {1\ov v_q^{\alpha_q^2}}}\right)
{\omega^3\ov \xi^4}\left[{1-{\pi\ov 4}{\omega\ov\xi}
(r+1/2)\sum_q{\alpha_q^2\ov v_q}+{\cal O}(\omega^2)}\right].
\label{eq:1/texp}
\eeqarr
If we retain only the lowest term in $\omega$ in this expansion, we note
that upon substituting this into Eq.~(\ref{eq:sigmazzwint}) we would
encounter the divergent sum
\beq
\label{eq:divsum}
\sum_{r=0}^{\8}(-1)^r.
\eeq
By considering the solvable point, where $v_q=v_0$ independent of $q$, it is
clear that the proper regularization of this sum is
\beq
\label{eq:regsum}
\sum_{r=0}^{\8}(-1)^r=\lim_{\delta\rightarrow 0}
\sum_{r=0}^{\8}(-e^{-\delta})^r=\lim_{\delta\rightarrow 0}{1\ov 1+e^{-\delta}}
={1\ov 2},
\eeq
and thus from Eqns.~(\ref{eq:sigmazzwint}), (\ref{eq:1/texp}), and
(\ref{eq:regsum}) we have
\beq
\label{eq:smallomegacond}
\Re e\, \ol{\sigma^{zz}}(\omega)
\stackrel{\omega\rightarrow 0}{\longrightarrow}
{\lambda^2 d\ov 48\pi^2}\left({\prod_q {1\ov
v_q^{\alpha_q^2}}}\right){\omega^2\ov \xi}.
\eeq
Comparing this with the small frequency asymptotics of
Eq.~(\ref{eq:sigmazzwell0}) at $T=0$:
\beq 
\label{eq:smallomegacond2}
\Re e\, \ol{\sigma^{zz}}(\omega)
\stackrel{\omega\rightarrow 0}{\longrightarrow}
{\lambda^2 d\ov 48\pi^2}{1\ov v_0^4}{\omega^2\ov \xi},
\eeq
we see that the interactions do not modify the result that the
conductivity goes to zero in the DC limit at zero temperature.  While
the power of $\omega$ is the same in Eqns.~(\ref{eq:smallomegacond})
and (\ref{eq:smallomegacond2}), the interactions do modify the
prefactor.  

To summarize, we have considered how interactions away from the
solvable point modify the vertical conductivity at $T=0$.  It should
be noted that in arriving at Eq.~(\ref{eq:sigmazzwint}) the
interactions were treated exactly.  We find that at large frequencies
the interactions soften the decay of the conductivity, but it remains
exponential, while at small frequencies the interactions do not change
the frequency exponent of the conductivity.  We remark that the
difficulty encountered when attempting to apply the method presented
in this section to investigate the effect of interactions at a finite
temperature is that in the finite-$T$ version of
Eq.~(\ref{eq:retardC}) there are complicated branch point
singularities on both sides of the real-$y$ axis, see
Eq.~(\ref{eq:IzIz}).  Nevertheless, we believe that even at a finite
temperature the asymptotics of $\sigma^{zz}(\omega)$ are not modified
by interactions.

\section{Irrelevant Tunneling}
\label{sec:irrelevant}

In the previous section we have been concerned exclusively with
the case of marginal tunneling, namely the 020 and 131 multilayer
states.  The calculation presented in Section~\ref{ss:dirty-marginal} can
readily be extended to the more general case of the $nmn$ state.
Essentially the only change in the computation is that the scaling
dimension of the tunneling operator, $\delta_{\lambda}=m-n$, differs
from its marginal value of 2.  Therefore, the poles in the complex
$y$-plane for the integrand in Eq.~(\ref{eq:intsinh}) are now of order
$2(m-n)$.  One finds that for the solvable point $g=nv/m$:
\beq
\label{eq:generalsigma}
\Re e\,\ol{\sigma^{zz}_{nmn}}(\omega)={\lambda^2 d\ov
(2\pi)^{2\delta_{\lambda}-1}v_0^{2\delta_{\lambda}}}{\Wt(\omega/v_0) \ov
(2\delta_{\lambda}-1)!}\prod_{k=1}^{\delta_{\lambda}-1}
[\omega^2+(2\pi kT)^2].  
\eeq

For the case of relevant tunneling, $\delta_{\lambda}=m-n=1$, the
product over $k$ is absent in Eq.~(\ref{eq:generalsigma}), and if we
use $\Wt(\omega/v_0)=(2\Delta)/(\omega^2+(\Delta/v_0)^2)$ appropriate
for delta-correlated disorder we reproduce the behavior of the chiral metal.

In the remaining cases, $\delta_{\lambda}\geq 2$, we see at $T=0$ the
conductivity vanishes as $\omega\rightarrow 0$ as
$\omega^{2(\delta_{\lambda}-1)}$, and at $\omega=0$ it vanishes with
temperature as $T^{2(\delta_{\lambda}-1)}$.  In the absence of
interlayer correlations, \ie for the $0m0$ state, this means the DC
conductivity obeys $\sigma^{zz}\sim T^{2m-2}$.  This disagrees with
the result of Balents and Fisher, who find by a scaling argument that
$\sigma^{zz}\sim T^{2m-3}$\cite{Balents-Fisher}.  While it is unclear
whether Balents and Fisher have in mind the clean system or the
disordered system, we believe the reconciliation of this discrepancy is
that, as we found for the marginal case, the temperature scaling of
the DC conductivity is different in the clean and disordered systems.
If we consider the clean limit of Eq.~(\ref{eq:generalsigma}) by
writing $\Wt(\omega/v_0)=(v_0/T)\delta(\omega/T)$, and then conjecture
that the interactions broaden the delta function, we find:
\beq
\label{eq:broadendelta2}
\Re e\,\ol{\sigma^{zz}_{nmn}}(\omega)\propto {\lambda^2 d\ov T}{g\ov 
(\omega/T)^2+{\cal O}(g^2)}
\prod_{k=1}^{\delta_{\lambda}-1}[\omega^2+(2\pi kT)^2],
\eeq
where $g$ is the dimensionless interaction strength.  The broadening
of the delta function by interactions is suggested by the fact that
the nearest-neighbor density-density interaction Hamiltonian does not
commute with the current operator $I^z$.  For the case of the $0m0$
state the DC conductivity of Eq.~(\ref{eq:broadendelta2}) goes to
zero as $T^{2m-3}$, in agreement with Balents and Fisher. 

Finally, returning to the disordered case (\ref{eq:generalsigma}), we
see that for the 050 state the DC conductivity scales as $T^8$, while
for 131 it scales as $T^2$ (\ref{eq:sigmazzmarg}).  Recall from
Eq.~(\ref{eq:nu_nmn}) that both of these states occur at the same
electron density, $\nu=1/5$ per layer.  In their numerical work, QJM
found a phase transition between these states as the layer separation
$d$ was varied.  The significant difference in the temperature scaling
of the DC vertical conductivity in these states suggests a way to
experimentally detect this phase transition.

\section{Fermionization}
\label{sec:fermionizations}

In certain special cases the multilayer edge theory can be solved
exactly.  The case of $N=2$ layers is discussed extensively
elsewhere\cite{nps1,nps2}, and examples of exact solutions for
$N=4$ layers are given in Appendix~\ref{sec:exactN=4}.  A potentially useful
first step to solving the edge theory for arbitrary $N$ is to find a
fermionic representation.  In this section we discuss the
fermionization of the edge theory for an arbitrary number of layers
$N$.  In some cases the procedure involves the introduction of
auxiliary degrees of freedom as discussed in Ref.~\cite{nps1}.  

We begin with the observation that the scaling
dimension~(\ref{eq:deltalambda}) of the tunneling operator is always
an integer.  This underlies the fact that the chiral
Hamiltonian~(\ref{eq:multiHfull}) can be fermionized exactly, with the
tunneling term expressed as a product of Fermi operators
whose number equals twice the scaling dimension.

The usual mapping between chiral radius-one bosons and fermions
requires {\em independent\/} bosonic fields.  Even though we are
interested in correlated states with non-trivial
commutators~(\ref{eq:multiCR}), these can be readily diagonalized by a
linear transformation.  For example, for the 131 state we can write
\begin{equation}
  \label{eq:131-decomposition}
  u_j=\varphi_{j-1/2}+\varphi_j+\varphi_{j+1/2}, \quad j=1,2,\ldots, N
\end{equation}
where the $2N$ bosons $\varphi_j$ with integer and half-integer
indices have the usual commutation relations
$[\varphi_i(x),\,\varphi_j(x')]=i\pi\,\delta_{ij}\,\sgn(x-x')$.
Clearly, Eq.~(\ref{eq:131-decomposition}) is not a canonical
transformation because the number of degrees of freedom has been
doubled.  This formal difficulty can be circumvented if we introduce
an independent set of auxiliary bosons with half-integer indices
$\tilde u_{i+1/2}$, $i=1,2,\ldots,N$ with identical commutation
relations,
\begin{equation}
  \label{eq:aux-commutator}
  [\tilde u_{i+1/2}(x),\,\tilde
  u_{j+1/2}(x')]=i\pi\,K_{ij}\,\sgn(x-x'), 
  \quad   [u_i(x),\,\tilde u_{j+1/2}(x')]=0.
\end{equation}
Then, we can further write for the 131 state
\begin{equation}
  \label{eq:131-aux-decomposition}
    \tilde u_{j+1/2}=-\varphi_{j}+\varphi_{j+1/2}-\varphi_{j+1}, 
\end{equation}
and the transformation $u_i,{\tilde u}_{i+1/2}\mapsto \varphi_j$
becomes canonical.

At the end of a calculation the auxiliary degrees of freedom need to
be projected out.  Such a projection has been explicitly carried out by us
in Ref.~\cite{nps1} for the simpler case of correlated quantum Hall
bilayers.  However, since the auxiliary degrees of freedom are
independent of the physical ones, all quantum-mechanical averages
involving physical quantities will automatically be correct,
independent of the Hamiltonian chosen for the additional bosons
$\tilde u_{i+1/2}$.  In the following we select this Hamiltonian in
the form~(\ref{eq:multiH0}), \ie  identical to that for the physical
bosonic fields sans the tunneling term.  In this case, the
transformations~(\ref{eq:131-decomposition}) and
(\ref{eq:131-aux-decomposition}) give
\begin{eqnarray*}
  {\cal H}_{131}&=&\ix \Biggl\{{1\over{4\pi}}\,\sum_{i,j=1}^N 
  \tilde V_{ij}\,\left(:\!\delx \varphi_i\, \delx \varphi_j\!:+     
    :\!\delx \varphi_{i+1/2}\, \delx \varphi_{j+1/2}\!:\right)\\ 
  &\relax&+ \sum_j \left[{\lambda\over(2\pi a)^2}
      \,  :\!e^{\textstyle i(\varphi_{j-1/2}+ \varphi_j- \varphi_{j+1}
      - \varphi_{j+3/2})}\!:+\,\,{\rm h.c.}\right]\Biggr\},
\end{eqnarray*}
where the modified interaction matrix is given by the matrix product
$\tilde V_{ij}\equiv K_{il} V_{lj}$.  Now we can use the
standard fermionization prescription:
\beq
\label{eq:SFP}
\psi_{j+\alpha}={1\ov\sqrt{2\pi a}}e^{i\varphi_{j+\alpha}},
\eeq
where $\alpha=0,1/2$.  In particular, in the fermionic representation
the original electron annihilation operator is written as a product of
{\em three\/} fermion operators, $\Psi_j
\propto \psi_{j-1/2}\psi_{j}\psi_{j+1/2}$, while the complete fermionized
Hamiltonian for the surface of the 131 system becomes
\begin{eqnarray}
    {\cal H}_{131}&=&\ix\biggl\{-\sum_{j,\alpha}
    {\tilde V}_{jj}:\!\psi_{j+\alpha}^\dagger i\delx
    \psi_{j+\alpha}\!:
    +{1\over2}\sum_{\alpha,i\neq j}\tilde V_{ij}
    :\!\psi_{i+\alpha}^\dagger \psi_{i+\alpha}\!:
    \,:\!\psi_{j+\alpha}^\dagger 
    \psi_{j+\alpha}\!:
    \nonumber\\
    & &    +\sum_j\left[\lambda\, \psi_{j-1/2}^\dagger \psi_{j}^\dagger
    \,\psi_{j+1} 
    \psi_{j+3/2}+{\rm h.c.}\right]\biggr\}.
  \label{eq:fermionized-131}  
\end{eqnarray}
A curious feature of this transformation is that the introduced
auxiliary degrees of freedom can be thought of as located at the
layers sandwiched {\em between\/} the physical layers---this is the
reason for assigning half-integer indices to them.  

Similar fermion representations also exist for other correlated
multilayer states.  For the bosonic state 020, the interlayer correlations
are absent, and the commutation relations can be diagonalized by the
transformation
\begin{equation}
  \label{eq:020-decomposition}
  u_j=\varphi_{j\uparrow}+\varphi_{j\downarrow}, 
\end{equation}
where the auxiliary degrees of freedom labeled by the pseudospin index
are located in the same layers.  The fundamental boson operator can be
written as $\Psi_j=\psi_{j\uparrow}\psi_{j\downarrow}$, and the full
Hamiltonian is
\begin{eqnarray}
    {\cal H}_{020}&=&\ix\biggl\{-\sum_{j,\sigma}{\tilde V}_{jj}
    :\!\psi_{j\sigma}^\dagger i\delx
    \psi_{j\sigma}\!:
    +{1\over2}\sum_{\sigma,i\neq j}\tilde V_{ij}
    :\!\psi_{i\sigma}^\dagger \psi_{i\sigma}\!:
    \,:\!\psi_{j\sigma}^\dagger 
    \psi_{j\sigma}\!:
    \nonumber\\
    & &    +\sum_j\left[\lambda\, \psi_{j\uparrow}^\dagger
    \psi_{j\downarrow}^\dagger 
    \,\psi_{j+1\downarrow} 
    \psi_{j+1\uparrow}+{\rm h.c.}\right]\biggr\},
    \quad\sigma=\uparrow,\downarrow. 
  \label{eq:fermionized-020}   
\end{eqnarray}
Despite the similarity with its counterpart for the 131 state
[Eq.~(\ref{eq:fermionized-131})], this Hamiltonian involves subtly
different correlations.  

In the case of 020 state, one can also avoid doubling the number of
degrees of freedom by using the alternative fermionization introduced
in Ref.~\cite{nps1}.  Specifically, the neighboring layers are
combined into pairs,
\begin{equation}
  u_{2n+1}=\varphi_{2n+1}+\varphi_{2n+2}, \quad u_{2n+2}=
  \varphi_{2n+1}-\varphi_{2n+2}, 
  \label{eq:fermionization-020-alt-transf}
\end{equation}
and the tunneling operators are fermionized alternatingly as anomalous
two- or four-fermion combinations,
\begin{equation}
  \label{eq:fermionized-020-tunnel}
  {1\ov (2\pi a)^{2}}e^{i(u_1-u_2)}=\psi_2^\dagger i\partial_x
  \psi_2^\dagger, \quad 
  {1\ov (2\pi a)^{2}}\,e^{i(u_2-u_3)}=\psi_1^\dagger \psi_2
  \psi_3\psi_4,\ldots, 
\end{equation}
where $\psi_j$ is given in Eq.~(\ref{eq:SFP}).  In cases where the
number of layers $N$ is small, \eg $N=2,3,4$, alternative
fermionizations exists for the 020 and 131 states, which in some cases
allow exact solutions, see Ref.~\cite{nps1} and
Appendix~\ref{sec:exactN=4}.

The fermionization is also very simple for the pyramid states.  As an
example consider the 12321 state where the commutation relations are
satisfied if we write
\begin{equation}
  u_j= \varphi_{j-1}+ \varphi_{j}+ \varphi_{j+1}, 
  \label{eq:fermionization-12321-transf}
\end{equation}
and the corresponding tunneling operator is bilinear, 
\begin{equation}
  \label{eq:fermionized-12321-tunnel}
  {1\ov 2\pi a}e^{i(u_j-u_{j+1})}=\psi_{j-1}^\dagger \psi_{j+1}.
\end{equation}
The edge theory with tunneling is quadratic and therefore exactly
soluble, except for the complication of the soft modes mentioned in
Section~\ref{sec:edge}.

\section{Summary}
We have investigated the fractional quantum Hall effect in infinite
layer systems.  In the bulk we find quasiparticles with several unusual
(irrational) features and at the edge we find surface phases which can
be considered ``chiral semi-metals'' when the interlayer tunneling is
not relevant.  In addition we have presented fermionizations of the
edge theory which may prove useful in finding exact solutions.

\section{Acknowledgments}

We would like to acknowledge support by an NSF Graduate Research
Fellowship (JDN), DOE Grant DE-FG02-90ER40542 (LPP) as well as NSF
grant No.\ DMR-99-78074, US-Israel BSF grant No.\ 9600294, and
fellowships from the A.~P.~Sloan Foundation and the David and Lucille
Packard Foundation (SLS).  LPP and SLS thank the Aspen Center for
Physics for its hospitality during the completion of part of this
work.

\appendix
\section{Model problem with marginal tunneling}
\label{sec:mmmm}

Here we consider a model problem with marginal interlayer tunneling.
The Hamiltonian is chosen in analogy with the
Hamiltonian~(\ref{eq:multiHwdirt2}), 
\begin{equation}
  \label{eq:m4-initial-Hamiltonian}
  {\cal H}=\ix
  \left\{v\psi_j^\dagger
  i\partial_x\psi_j+{1\over2}\left[\lambda\,e^{i\gamma_{j}(x)}
  \left(\psi_j^\dagger
  i\partial_x\psi_{j+1}-i\partial_x\psi_j^\dagger\,\psi_{j+1}\right)+{\rm
  h.c.}\right]  \right\}, 
\end{equation}
with the marginal tunneling operator in parenthesis of a form
reminiscent of the two-fermion-fermionized version of the marginal
tunneling operator~(\ref{eq:fermionized-020-tunnel}).  The great
advantage of the Hamiltonian~(\ref{eq:m4-initial-Hamiltonian}) is its
linearity, which allows us to calculate the conductivity for this
model exactly.  In order to do this, we perform a gauge transformation
$\psi_j\to e^{i\zeta_j(x)}\psi_j$, with
$\zeta_{j+1}-\zeta_j\equiv\gamma_j$.  This gives, in obvious matrix
notation,
\begin{equation}
  \label{eq:m4-gauged-Hamiltonian}
  {\cal H}=\ix
  \left[\Psi^\dagger \Lambda
  i\partial_x\Psi+{1\over2}
  \Psi^\dagger   \left(\Lambda \hat V+\hat V\Lambda\right)
    \Psi \right], 
\end{equation}
where $\Psi\equiv (\psi_1\,\psi_2\,\ldots)^{T}$, the tri-diagonal matrix
$\Lambda_{ll'}\equiv v\delta_{ll'}
+\lambda(\delta_{l,l'+1}+\delta_{l+1,l'})$ is coordinate-independent,
while $\hat V\equiv {\rm diag}(V_1, V_2, \ldots )$ with $V_j\equiv
\partial_x\zeta_j$.  The corresponding transverse current operator has
the form
\begin{equation}
  \label{eq:m4-current-z}
  I^z=\ix \Psi^\dagger \left[\hat t\,i\partial_x+{1\over2}\left(\hat
  t\hat V+\hat V \hat t\right)\right]\Psi,
\end{equation}
where $\hat t_{ij}\equiv i\lambda\,(\delta_{i,j+1}-\delta_{i+1,j})$.

The formal solution of this model (for {\em a given\/} realization of
disorder) can be written in the limit of infinite system size
($L\to\infty$) using the transfer matrix approach
described in Appendix~A of Ref.~\cite{nps2}.
Using the Kubo formula for the vertical conductance, we obtain
\begin{equation}
  \label{eq:m4-vertical-conductance-formula}
  G^{zz}(\omega)={L\ov
  Nd}\sigma^{zz}(\omega)={1\over2N\omega}\sum_k\ix\left(n_k-n_{k+\omega}\right){\rm
  Tr}\left[\tilde{\cal V}(x)S_k(x,y)\tilde{\cal V}(y)S_{k+\omega}(y,x)
  \right],
\end{equation}
where $n_k=(e^{\beta vk}+1)^{-1}$, the matrix in the vertex
\begin{equation}
  \label{eq:m4-vertex}
 \tilde{\cal V} (x)\equiv \Lambda^{-1/2} \left[\hat
  t\,i\partial_x+{1\over2}\left(\hat 
  t\hat V+\hat V \hat t\right)\right]\Lambda^{-1/2},
\end{equation}
and the transfer matrix 
\begin{equation}
  \label{eq:m4-transfer-matrix}
  S_k(a,b)=T_x\exp\left({-i\int_a^bdx\,\left[k\Lambda^{-1}
  -{1\over2}\Lambda^{-1/2}\left(\hat 
  t\hat V+\hat V \hat t\right)\Lambda^{-1/2}\right]}\right). 
\end{equation}

To compute the disorder averages we assume that $V_j(x)$ are independent
delta-correlated Gaussian random variables with zero average,
$\ol{V_i(x)\,V_{j}(y)}=\Delta\delta_{ij}\delta(x-y)$.
Then, a calculation at a finite temperature $T$ gives
\begin{equation}
  \label{eq:m4-answer}
  \ol{G^{zz}}={L\over 24}\left(\omega^2+4\pi^2
    T^2\right)\int_{-\pi}^\pi {dq\over2\pi} {t^2(q)\over\Lambda^3(q)}\,
  {2\tilde\Delta(q)\over\omega^2+\tilde\Delta^2(q)},
\end{equation}
where $\Lambda(q)=v+2\lambda\cos q$, $t(q)=2\lambda\sin q$, and
$$
\tilde\Delta(q)={\Delta\over4}\left[2{\Lambda(q)}+{1}
    +{\Lambda^2(q)\over\sqrt{v^2-\lambda^2}}\right]. 
$$

As in the physical marginal cases, 020 and 131, at zero temperature
and in the absence of disorder the conductivity vanishes.  In the
leading order in $\lambda$ Eq.~(\ref{eq:m4-answer}) is identical to
the result for 020 with delta-correlated disorder
(\ref{eq:sigmazzmarg2}).  However, in the clean limit we find
$G^{zz}\propto \delta(\omega)$ for the model problem, which is not the
correct behavior for the 020 and 131 states.  The difference arises
because, in contrast to 020 and 131, in the model problem
the current operator $I^z$ (\ref{eq:m4-current-z}) commutes with the
full Hamiltonian (\ref{eq:m4-gauged-Hamiltonian}).

\section{Exact Solutions for $N=4$}
\label{sec:exactN=4}

In this appendix we describe exact solutions for the spectrum of the
edge theory for the special case of $N=4$ layers with periodic
boundary conditions.  We begin with a discussion of the 010 state with
both tunneling and interactions and then consider the 020 state.  We
remark that similar solutions exist for the 010, 020, and 131 states in
$N=3$ layers.

\subsection{010 State}
\label{ss:010forN=4}
The Hamiltonian for the 010 state is
\beq
\label{eq:H010N=4}
{\cal H}_{010}^{N=4}=\ix\left[{{v\ov 4\pi}\,:\!(\delx u_i)^2\!:\,+
{g\ov 2\pi}\,:\!\delx u_i\delx u_{i+1}\!:\,+{\lambda\ov 2\pi a}
\left({e^{-i(u_i-u_{i+1})}+\,{\rm h.c.}}\right)}\right],
\eeq
where $u_5\equiv u_1$.  A straightforward fermionization
$\psi_i=e^{iu_i(x)}/\sqrt{2\pi a}$ would yield a quartic Hamiltonian
because of the density-density interaction term.
We thus perform the orthogonal transformation
$u_i(x)=\Lambda_{ij}\phi_j(x)$ where
\beq
\label{eq:Lambda}
\Lambda={1\ov 2}\pmatrix{1 & 1 & 1 & 1 \cr
1 & 1 & -1 & -1 \cr 1 & -1 & 1 & -1 \cr
1 & -1 & -1 & 1}.
\eeq
The Hamiltonian (\ref{eq:H010N=4}) then reads
\beqarr
{\cal H}_{010}^{N=4}&=&\ix \left[{{1\ov
4\pi}v_i:\!(\delx\phi_i)^2\!:\,
}\right.\nonumber\\
&&\left.{{}+{\lambda \ov 2\pi a}\left({
e^{-i(\phi_2-\phi_3)}+e^{-i(\phi_3-\phi_4)}+
e^{i(\phi_2+\phi_3)}+e^{-i(\phi_3+\phi_4)}+\,
{\rm h.c}}\right)}\right],\label{eq:H010N=4.2}
\eeqarr
where $v_{1,3}=v\pm 2g$ and $v_2=v_4=v$.  The original fields,
$u_i$, are all radius $R=1$ chiral bosons.  We restrict
ourselves to the sector where the total topological charge of the
fields $u_i$ is constant, which corresponds to the conservation of
the total fermion number in the fermionic form of (\ref{eq:H010N=4}).
Then the requirement that the topological charges of the original
($u_i$) and transformed ($\phi_i$) bosons be integral is
consistent with taking the fields $\phi_i$ to all have unit radius
as well.  We can therefore fermionize according to
$\eta_i(x)=e^{i\phi_i(x)}/\sqrt{2\pi a}$ and then express the
fermions in terms of their Fourier components
$\eta_i(x)=\sum_k e^{ikx}c_{ik}/\sqrt{L}$.  In terms of these mode
operators the Hamiltonian (\ref{eq:H010N=4.2}) is
\beq
\label{eq:H010N=4.3}
{\cal H}_{010}^{N=4}=\sum_k\left[{v_ikc_{ik}^{\dag}c_{ik}
+\lambda\left({c_{2k}^{\dag}c_{3k}+c_{3k}^{\dag}c_{4k}
+c_{2k}c_{3-k}+c_{3k}^{\dag}c_{4-k}^{\dag}+\,{\rm h.c.}
}\right)}\right],
\eeq
where the momentum $k$ takes values in $(2\pi/L)(\Zint+1/2)$.
This form of the edge theory is quadratic and hence exactly soluble
via a Bogoliubov transformation.  We find:
\beq
\label{eq:H010N=4.4}
{\cal H}_{010}^{N=4}=\sum_k \epsilon_i(k)b_{ik}^{\dag}b_{ik},
\eeq
where $\{b_{ik},b^{\dag}_{jk'}\}=\delta_{ij}\delta_{kk'}$ are four
independent branches of fermionic oscillators with dispersions
\beq
\matrix{
\epsilon_1(k)=(v+2g)k,& 
\epsilon_3(k)=(v-g)k-\sqrt{(gk)^2+4\lambda^2},  \cr
\epsilon_2(k)=vk, &
\epsilon_4(k)=(v-g)k+\sqrt{(gk)^2+4\lambda^2} \cr}.
\label{eq:dispN=4}
\eeq
We find that two of the branches develop curvature if and only if
both $g$ and $\lambda$ are nonzero.

\subsection{020 State}
\label{ss:020forN=4}
Now consider the 020 state whose bosonic Hamiltonian is given above in
Eq.~(\ref{eq:H020}).  The fields $u_i$ all have unit radius, but
their commutators are not conventionally normalized, see
Eq.~(\ref{eq:multiCR}).  From our solution above for the 010 state we know
that we can perform an orthogonal transformation (\ref{eq:Lambda}) and
still have four radius one bosons $\phi_i$.  If we then define rescaled
fields $\theta_i(x)\equiv \phi_i(x)/\sqrt{2}$ the 020 Hamiltonian
becomes
\beqarr
{\cal H}_{020}^{N=4}&=&\ix\left[{{v_i\ov 8\pi}\,:\!(\delx\theta_i)^2\!:\,
}\right.\nonumber\\
&&\left.{{}+{\lambda\ov (2\pi a)^2}\left({
e^{-i\sqrt{2}(\theta_2-\theta_3)}+e^{-i\sqrt{2}(\theta_3-\theta_4)}+
e^{i\sqrt{2}(\theta_2+\theta_3)}+e^{-i\sqrt{2}(\theta_3+\theta_4)}+\,
{\rm h.c.}}\right)}\right],\label{eq:H020N=4}
\eeqarr
where $\theta_i$ are radius $R=1/\sqrt{2}$ chiral
bosons with conventionally normalized commutators, and the velocities
$v_i$ are the same as those given above in the solution of the 010
state. We can define a quartet of $\widehat{su}(2)_1$ KM
currents as in Eq.~(\ref{eq:multisu2}).  The Hamiltonian becomes
\beq
\label{eq:H020N=4.2}
{\cal H}_{020}^{N=4}=\ix\left[{{\pi v_i\ov 3}\,:\![{\bf J}_i(x)]^2\!:\,+
4\lambda\left({J^x_2J^x_3+J^x_3J^x_4}\right)
}\right],
\eeq
where we have again used the identity (\ref{eq:JztoJagain}).  Next, we
define a new set of currents $\tilde{J}^a_i(x)=R^{ab}J^b_i(x)$, which
also obey independent $\widehat{su}(2)_1$ algebras, where
$R^{ab}\equiv \delta^{az}\delta^{bx}+\delta^{ay}\delta^{by}
-\delta^{ax}\delta^{bz}$.  This rotation leaves the first term in the
Hamiltonian (\ref{eq:H020N=4.2}) unchanged and in the second term
gives $J_i^x\mapsto -\tilde{J}_i^z$.  If we express the rotated KM
currents, $\tilde{J}^a_i$, in terms of four new radius $R=1/\sqrt{2}$
chiral bosons $\tilde{\theta_i}$ via Eq.~(\ref{eq:multisu2}), we find
\beq
\label{eq:H020N=4.3}
{\cal H}_{020}^{N=4}=\ix{1\ov 4\pi}M_{ij}\,:\!\delx\tilde{\theta}_i
\delx\tilde{\theta}_j\!:\,,
\eeq
where
\beq
\label{eq:M}
M\equiv\pmatrix{v/2+g & 0 & 0 & 0 \cr
0 & v/2 & \lambda/\pi & 0 \cr
0 & \lambda/\pi & v/2-g & \lambda/\pi \cr
0 & 0 & \lambda/\pi & v/2 \cr}.
\eeq
We have succeeded in finding a quadratic representation of the
Hamiltonian which can be readily diagonalized by performing an
orthogonal transformation from the fields $\tilde{\theta_i}$
to new fields $\vartheta_i$.  The final form of the Hamiltonian is
\beq
\label{eq:H020N=4.4}
{\cal H}_{020}^{N=4}=\ix{1\ov 4\pi}\tilde{v}_i\,:\!
(\delx\vartheta_i)^2\!:\,,
\eeq
where the velocities are
\beq
\label{eq:finalvelocities}
\matrix{ \tilde{v}_1 = v/2+g & 
\tilde{v}_3=v/2-g/2 + \sqrt{(g/2)^2+2(\lambda/\pi)^2} \cr
\tilde{v}_2=v/2 &
\tilde{v}_4=v/2-g/2 - \sqrt{(g/2)^2+2(\lambda/\pi)^2} \cr}.
\eeq
The exact spectrum of the 020 state contains four independent free chiral
bosons whose velocities are renormalized by both interactions and
tunneling.

\section{Klein Factors}
\label{sec:Kleins}

In this appendix we discuss Klein factors which are needed to produce
the proper anticommutation relations between different species of
fermions.  One place where Klein factors are needed is in the
definition of the physical electron operators at the edges of 
each layer.  We modify the definition of the electron
operator at the edge of layer $i$ (\ref{eq:multirho,psi}) to read
\beq
\label{eq:newPsi}
\Psi_i(x)\propto e^{iA_{ij}N_j}e^{iu_i(x)},
\eeq
where $A$ is an $N\times N$ matrix and $N_j$ is the topological charge
of the Bose field $u_j(x)$ defined by
\beq
\label{eq:topcharge}
N_j={1\ov 2\pi}\ix \delx u_j(x).
\eeq
Using the commutation relations of the chiral bosons (\ref{eq:multiCR}) one
can readily show 
\beq
\label{eq:interPsi}
\Psi_i(x)\Psi_j(x')=\Psi_j(x')\Psi_i(x)e^{-i\pi K_{ij}\sgn(x-x')}
e^{-i(A_{ik}K_{kj}-A_{jk}K_{ki})},
\eeq
and the combination appearing in the tunneling operator (\ref{eq:tunnelop})
can be written
\beq
\label{eq:tunopwK}
\Psi_i(x)\Psi_{i+1}^{\dag}(x)\propto e^{i(A_{ik}-A_{i+1,k})N_k}
e^{i[u_i(x)-u_{i+1}(x)]}e^{i(A_{i+1,k}K_{k,i+1}-A_{i+1,k}K_{ki})}.
\eeq
If we write $A=\pi BK^{-1}$, for some matrix $B$, then we have
\beqarr
\Psi_i(x)\Psi_j(x')&=&\Psi_j(x')\Psi_i(x)e^{\mp i\pi
K_{ij}}e^{-i\pi(B_{ij}-B_{ji})},
\label{eq:BAC}\\
\Psi_i(x)\Psi_{i+1}^{\dag}(x)&\propto & e^{i[u_i(x)-u_{i+1}(x)]}
e^{i\pi(B_{i+1,i+1}-B_{i+1,i})}e^{i\pi(B_{il}-B_{i+1,l})(K^{-1})_{lk}N_k}
\label{eq:BTO}.
\eeqarr

From Eq.~(\ref{eq:BAC}) we see that for the electron operators in
different layers to anticommute we must have $(B_{ij}-B_{ji})\in
\Zint_{\rm even}$ if $K_{ij}$ is odd and $(B_{ij}-B_{ji})\in
\Zint_{\rm odd}$ if $K_{ij}$ is even.  From Eq.~(\ref{eq:BTO}) we see
that if the elements of the matrix $B$ are integers and
$(B_{il}-B_{i+1,l})(K^{-1})_{lk}$ is an integer for all $i$ and $k$,
then the tunneling operator will be the same as in the absence of the
Klein factors up to a sign which depends on the topological charge
sector.

As a concrete example, consider the multilayer 131 state.  If the number of
layers $N$ is such that $\det(K)$ is odd (\ie $N \,{\rm mod}\,3 \neq 2$)
then the choice
\beq
\label{eq:B}
B_{ij}=\left\{{\matrix{\det(K) & {\rm for}\,\, j-i\geq 2 \cr
0 & {\rm otherwise}}}\right.
\eeq 
gives $\{\Psi_i(x),\Psi_j(x')\}=0$ for all $i,j$ and modifies the
tunneling term by a factor of $\pm 1$.  Note that in correlation
functions, such as the current two-point function used to compute the
conductivity (\ref{eq:margC}), the tunneling operator is always
accompanied by its hermitian conjugate and therefore factors such as
$e^{i\pi(B_{il}-B_{i+1,l})(K^{-1})_{lk}N_k}$ will cancel.

In the above discussion we have constructed Klein factors out of the
topological charge operators of the Bose fields in order to give the
correct anticommutation relations between the physical electron
operators without significantly modifying the tunneling Hamiltonian.
When considering the fermionization of the multilayer edge theory, as
discussed in Section~\ref{sec:fermionizations}, we have a different
goal.  In addition to ensuring that the electron operators anticommute
we want the different species in the fermionized Hamiltonian to obey
proper anticommutation relations.  In this case an alternative
approach to the one described above proves advantageous.  This
procedure is best described by considering a specific case, for
example the 131 state.  One can introduce $2N$ unitary operators
$F_{i+\alpha}$, where $\alpha=0,1/2$ and $i=1,\ldots,N$, by formally
enlarging the Hilbert space.  These operators commute with the Bose
fields $u_i$ and obey \beq
\label{eq:Falgebra}
\{F_{i+\alpha},F_{j+\alpha'}\}= 
\{F^{\dag}_{i+\alpha},F^{\dag}_{j+\alpha'}\}= 
\{F_{i+\alpha},F^{\dag}_{j+\alpha'}\}=0,
\eeq
for $i+\alpha\neq j+\alpha'$.  The electron operators are modified
according to
\beq
\label{eq:Psiw/F}
\Psi_i(x)\propto F_{i-1/2}F_{i}F_{i+1/2}e^{iu_i(x)}.
\eeq
It is a straightforward exercise to demonstrate that
Eqns.~(\ref{eq:Falgebra}) and (\ref{eq:Psiw/F}) imply that electron
operators in different layers properly anticommute.  The tunneling
operator is now
\beq
\label{eq:tunopw/F}
\Psi_j(x)\Psi_{j+1}^{\dag}(x)\propto e^{i[u_j(x)-u_{j+1}(x)]}
F_{j-1/2}F_jF^{\dag}_{j+1}F^{\dag}_{j+3/2}.
\eeq
If we then follow the procedure described in
Section~\ref{sec:fermionizations} with the modification that the new
fermion species each absorb a factor of $F$:
\beq
\label{eq:psiw/F}
\psi_{j+\alpha}\equiv {1\ov \sqrt{2\pi a}}
F_{j+\alpha}e^{i\varphi_{j+\alpha}(x)},
\eeq
we arrive at the same fermionized Hamiltonian
(\ref{eq:fermionized-131}), but now with the different fermion species
properly anticommuting.

\section{Proof that $\Re \lowercase{e} \,\sigma^{\lowercase{zz}}
(\omega)\propto \delta(\omega) \leftrightarrow [I^{\lowercase{z}},H]=0$}
\label{sec:converse}

In this appendix we prove that the real part of the $z$-axis
conductivity is proportional to a delta function in frequency if and
only if the current operator commutes with the full Hamiltonian of the
edge theory.  The basic connection is of course the Kubo formula
(\ref{eq:Kubo2}):
\beq
\label{eq:moreKubo}
\Re e\, \sigma^{zz}(\omega)=
{1\ov 2\omega}{1\ov NLd}\int_{-\8}^{\8}dt\, e^{i\omega t}
\la [I^z(t),I^z(0)]\ra.
\eeq
One of the directions is trivial: if $[I^z,H]=0$, then $I^z(t)$
is independent of time, and hence the commutator in Eq.~(\ref{eq:moreKubo})
vanishes identically implying that for all $\omega\neq 0$, 
$\Re e\,\sigma^{zz}(\omega)=0$ and therefore $\Re e\,\sigma^{zz}\propto
\delta(\omega)$.   

We now prove the converse, \ie we assume 
$\Re e\,\sigma^{zz}\propto\delta(\omega)$.  Inverting the Fourier transform
in Eq.~(\ref{eq:moreKubo}) we find
\beq
\label{eq:invertKubo}
\la[I^z(t),I^z(0)]\ra=2NLd\int_{-\8}^{\8}d\omega \,e^{-i\omega t}
\omega \Re e\,\sigma^{zz}(\omega).
\eeq
Using our assumption in this equation gives
\beq
\label{eq:IIiszero}
\la[I^z(t),I^z(0)]\ra=\sum_n e^{-\beta E_n}
\la n|[I^z(t),I^z(0)]|n\ra=0,
\eeq
where $|n\ra$ and $E_n$ are
eigenstates and eigenvalues of $H$.  Note that the sum in
Eq.~(\ref{eq:IIiszero}) is a linear combination of terms $\la n|
[I^z(t),I^z(0)]|n\ra$ with non-negative coefficients which depend on
$\beta$.  Since the sum must vanish for arbitrary $\beta$ we conclude that
\beq
\label{eq:eachterm}
\la n|[I^z(t),I^z(0)]|n\ra=0,\quad \forall\,\, n.
\eeq
Writing $I^z(t)=\exp(iHt)I^z(0)\exp(-iHt)$ and inserting a summation
over a complete set of states $\{|m\ra\}$, Eq.~(\ref{eq:eachterm}) is
equivalent to
\beq
\label{eq:sumonm}
\sum_m \sin[(E_n-E_m)t]|\la n|I^z(0)|m\ra|^2=0,\quad \forall\,\, n,\quad
\forall\,\, t.
\eeq
The quantities $|\la n|I^z(0)|m\ra|^2$ are clearly non-negative.  Since the
sum in Eq.~(\ref{eq:sumonm}) must vanish for all $n$ and at all times
$t$ we can conclude
\beq
\label{eq:I^2is0}
|\la n|I^z(0)|m\ra|^2=0, \quad \forall\,\, n\neq m.
\eeq
This equation says that the current has no off-diagonal matrix elements in
the basis of energy eigenstates, \ie the current operator $I^z$ is
diagonal in the basis of eigenstates of $H$.  Since this means $I^z$
and $H$ can be simultaneously diagonalized it implies that they commute:
$[I^z(0),H]=0$, completing the proof.


\begin{thebibliography}{99}

\bibitem{stormer1}
H. L. Stormer, J. P. Eisenstein, A. C. Gossard, W. Wiegmann, and
K. Baldwin, Phys.\ Rev.\ Lett.\ {\bf 56}, 85 (1986).

\bibitem{organics}
S. Uji, C. Terakura, M. Takashita, T. Terashima, H. Aoki,
J. S. Brooks, S. Tanaka, S. Maki, J. Yamada, S. Nakatsuji, Phys.\
Rev.\ B {\bf 60}, 1650 (1999), and references therein.

\bibitem{Chalker-Dohmen}
J. T. Chalker and A. Dohmen, Phys.\ Rev.\ Lett.\ {\bf 75}, 4496 (1995).

\bibitem{Balents-Fisher}
L. Balents and M. P. A. Fisher, Phys.\ Rev.\ Lett.\ {\bf 76}, 2782 (1996).

\bibitem{Betouras-Chalker}
J. J. Betouras and J. T. Chalker, {\tt cond-mat/0005151}, and
references therein.

\bibitem{druist}
D. P. Druist, P. J. Turley, K. D. Maranowski, E. G. Gwinn, and
A. C. Gossard, Phys.\ Rev.\ Lett.\ {\bf 80}, 365 (1998).

\bibitem{brey}
L. Brey, Phys.\ Rev.\ Lett.\ {\bf 81}, 4692 (1998). 

\bibitem{gm1}
S.~M. Girvin and A.~H. Mac{D}onald, {\em in} Novel Quantum Liquids in
Low-Dimensional Semiconductor Structures, ed. S. DasSarma and
A. Pinczuk (Wiley, New York, 1995).

\bibitem{QJM}
X. Qiu, R. Joynt, and A. H. MacDonald, Phys.\ Rev.\ B {\bf 40}, 11943
(1989); Phys.\ Rev.\ B {\bf 42}, 1339 (1990).

\bibitem{nps3}
J. D. Naud, L. P. Pryadko, and S. L. Sondhi, {\tt cond-mat/0006432}.
  
\bibitem{nps1}
J. D. Naud, L. P. Pryadko, and S. L. Sondhi, Nucl.\ Phys.\ B 
{\bf 565}, 572 (2000).

\bibitem{nps2}
J. D. Naud, L. P. Pryadko, and S. L. Sondhi, unpublished,
{\tt cond-mat/0006173}.

\bibitem{greiter-mcdonald}
M. Greiter, and I. A. McDonald, Nucl.\ Phys.\ B {\bf 410}, 521 (1993).

\bibitem{chen}
Y. J. Chen, J.\ Phys.\ Condens. Matter {\bf 10}, 2437 (1998).

\bibitem{ASW}
D. Arovas, J. R. Schrieffer, and F. Wilczek, Phys.\ Rev.\ Lett.\
{\bf 53}, 722 (1984).

\bibitem{wen}
X.-G. Wen and A. Zee, Phys.\ Rev.\ B {\bf 46}, 2290 (1992).

\bibitem{karlhede-sondhi}
A. Karlhede, K. Lejnell, and S. L. Sondhi, Phys.\ Rev.\ B {\bf 60},
15948 (1999).

\bibitem{fn-QJM}
In their Monte Carlo calculations QJM found that for some values of the
layer separation $d$ the ground state had a non-zero value for the
next-nearest-neighbor correlation exponent\cite{QJM}.

\bibitem{mahan}
G. D. Mahan, Many Particle Physics, 2nd ed. (Plenum, New York, 1990).

\end{thebibliography}
\end{document}